\documentclass[aps,10pt,twocolumn,superscriptaddress,preprintnumbers,floatfix,pre]{revtex4-1}
\usepackage{mathrsfs, hyperref}
\usepackage{amssymb, amsbsy, amsmath, latexsym, dsfont, array, layout, graphics,mathrsfs,braket,amsfonts,amsthm,
  amssymb,graphicx,subfigure, youngtab,color,bm,mathtools,verbatim,url,cleveref,natbib,hypernat,extarrows,inputenc,multirow,bbm,dsfont}
\usepackage{graphicx}
\usepackage{dcolumn}
\usepackage{bm}
\usepackage[dvipsnames]{xcolor}
\usepackage[normalem]{ulem}
\usepackage[T1]{fontenc}
\hypersetup{
     colorlinks = true,
     linkcolor = black,
     citecolor = black
}
\graphicspath{ {./images/} }
\newcommand{\ii}{\mathrm{i}}

\newcommand{\Arg}{\operatorname{Arg}}
\newcommand{\Ne}{N_e}
\newcommand{\Nphi}{N_\phi}

\newcommand{\bk}{\bm k}
\newcommand{\br}{\bm r}
\newcommand{\bp}{\bm p}
\newcommand{\bq}{\bm q}
\newcommand{\bG}{\bm G}

\newcommand{\be}{\bm\eta}

\newcommand{\ba}{\bm a}
\newcommand{\bb}{\bm b}
\newcommand{\bA}{\bm A}
\newcommand{\qell}{\ell_\eta}
\renewcommand{\ket}[1]{\lvert #1\rangle}

\newcommand{\mel}[3]{\langle #1\vert #2\vert #3\rangle}
\allowdisplaybreaks

\begin{document}
\title{Lattice-Induced Splitting, Connectivity, and Topology of Anyon Bands}
\author{Siddhartha Sarkar}
\email{sarkars@umd.edu}
\affiliation{Max Planck Institute for the Physics of Complex Systems, N\"othnitzer Stra\ss e 38, 01187 Dresden, Germany}
\affiliation{Condensed Matter Theory Center and Joint Quantum Institute, University of Maryland, College Park, MD 20742, USA}
\author{Xiaohan Wan}
\affiliation{%
Department of Physics, University of Washington, Seattle, WA 98195, USA}

\begin{abstract}
  Fractional Chern insulators realize fractional quantum Hall states without external magnetic fields, while their underlying lattice periodicity allows otherwise dispersionless anyons to acquire finite bandwidth. Here we show that flux--lattice commensurability can also give rise to multiple anyon bands with nontrivial topology and symmetry-protected connectivity. In multiband vortexable systems, commensurability between the lattice periodicity and the effective magnetic unit cell of fractionally charged quasiholes leads to the folding of a quasihole Landau level into multiple bands. Interactions lift their degeneracy, producing dispersive bands with Chern numbers distinct from that of the parent quasihole Landau level. We further demonstrate that these Chern numbers can be redistributed through Dirac crossings under continuous deformations that preserve exact electronic flatness, while crystalline symmetries can enforce protected band touchings. Our results establish that fractionalized excitations can exhibit the rich band topology familiar from electronic Bloch bands, opening possibilities for nontrivial anyon dynamics beyond conventional Landau-level physics.
\end{abstract}
\maketitle

\noindent\textit{Introduction.}---
The interplay of multiple Bloch bands and crystalline symmetries underlies much of the richness of electronic band theory, from band inversions and topological bands to symmetry- or topology-protected nodal points and lines~\cite{young2015dirac,bradlyn2017topological}. The emergence of topological flat bands has brought lattice periodicity and space-group symmetries into the realm of fractional quantum Hall (FQH) physics, whose elementary excitations are anyons. Can these crystalline ingredients endow anyons with equally rich band structures, including nontrivial topology and symmetry-protected crossings and connectivity, even when the underlying electronic bands remain exactly flat?

Fractional Chern insulators (FCIs) realize FQH states in topological lattice bands~\cite{tang2011high,sun2011nearly,neupert2011fractional,sheng2011fractional,Regnault2011fractional}, as recently demonstrated in moir\'e materials~\cite{cai2023signatures,zeng2023thermodynamic,park2023observation,xu2023observation,lu2024fractional,xie2025tunable}. Unlike the dispersionless quasiholes in conventional Landau levels (LLs), anyons in FCIs can acquire finite dispersion due to lattice periodicity. Recent studies have explored their dispersions and the roles of interactions and nonuniform quantum geometry~\cite{goncalves2025spinless,liu2025quasiparticles,schleith2025anyon,yan2025anyon,iyer2026dispersion}, as well as the topology of composite-fermion excitation bands~\cite{pu2026composite}. The possibility of mobile anyons has also motivated proposals for unconventional metallic and superconducting phases upon doping FCIs ~\cite{shi2025doping,nosov2026anyon,
lotrivc2026phases,fan2026hidden,
shi2026superconductivity}. These advances motivate exploring whether the crystalline structure of FCIs can give rise to an entire multiband theory of anyons, beyond simply introducing finite bandwidth.

To uncover the general principles governing anyon band structures, we consider flat Chern bands with ideal quantum geometry~\cite{wu2012fractional,parameswaran2013fractional,roy2014band,wu2015fractional,ledwith2020fractional,ledwith2021strong,wang2021exact,mera2021engineering,ledwith2023vortexability} satisfying $\operatorname{tr}g(\bk)=
|F_{xy}(\bk)|$, where $g$ and $F_{xy}$ are the quantum metric and Berry curvature, independent of their microscopic realization. Ideal bands with Chern number $|C|=1$ admit a mapping to the zero-mode manifold of a generalized Aharonov--Casher Hamiltonian, describing charged particles in an effective, spatially varying magnetic field $B(\bm r)$~\cite{aharonov1979ground,ledwith2020fractional,dong2022dirac,estienne2023ideal}. This mapping provides a unified classification based on the commensurability between magnetic flux and lattice periodicity~\cite{zak1964magnetic,dubrovin1980ground}. The average field $B_0=\langle B(\mathbf r)\rangle$ and the flux quantum $\Phi_0$ define the \emph{flux-quantum area} $A_\Phi=\Phi_0/|B_0|$, while the periodicity of $B(\mathbf r)$ defines a primitive unit-cell area $A_{\rm uc}$. For $A_{\rm uc}/A_\Phi=n/m$, with coprime positive integers $n$ and $m$, the smallest common supercell has area $A_{\rm M}=mA_{\rm uc}=nA_\Phi$ and encloses $n$ flux quanta. The familiar $n=m=1$ case is exemplified by the chiral limit of twisted bilayer graphene~\cite{ledwith2020fractional,wang2021exact} and adiabatic limit of twisted MoTe$_2$~\cite{morales2024magic,shi2024adiabatic}, yielding a single exactly flat Chern band per flavor. Beyond this well-studied limit, two qualitatively distinct regimes reveal new possibilities. For $n=1$, $m>1$, the rational flux per primitive cell resembles the Hofstadter problem ~\cite{hofstadter1976energy,
thouless1982quantized,dubrovin1980ground}, while the magnetic unit cell still supports a single exactly flat band. A fundamentally different regime emerges for $n>1$, where the common supercell encloses multiple flux quanta, folding a single LLL into $n$ exactly degenerate flat bands. Protected by the underlying zero-mode structure, these bands collectively retain ideal quantum geometry and total Chern number $C=-1$~\cite{wan2023topological,le2022double,becker2022fine,becker2023degenerate,sarkar2023symmetry,sarkar2025unconventional}. Such multiband systems already exhibit unconventional fractional states with a mismatch between filling per unit cell and Hall conductivity~\cite{sarkar2025unconventional}. As we will show below, this multiband structure also provides the key to unlocking rich anyon band topology and symmetry-protected phenomena, even in the absence of electronic dispersion.

\begin{figure}[!t]
\centering
\includegraphics[width=0.46\textwidth]{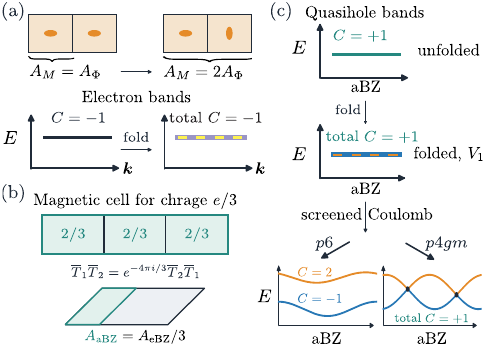}

\caption{
\textbf{Schematic of flux--lattice commensurability
and topological quasihole bands..} (a) A reference electronic Landau level (LL) with one flux quantum per unit cell of area $A_\Phi=2\pi\ell_B^2$ folds into two exactly degenerate flat bands when the moir\'e unit cell is enlarged to $A_{\rm M}=2A_\Phi$. The total Chern number of the two degenerate bands remains $C=-1$. (b) At Laughlin filling $\nu_{\rm LLL}=1/3$, corresponding to $\nu_{\rm M}=2/3$ electrons per moir\'e cell, a quasihole of charge $e/3$ experiences an effective flux of $2/3$ per moir\'e cell. Three moir\'e cells therefore enclose two quasihole flux quanta, defining a quasihole magnetic unit cell. The quasihole magnetic translation algebra is $\overline T_1\overline T_2
=e^{-4\pi\ii/3}\overline T_2\overline T_1$, and the magnetic anyon Brillouin zone (aBZ) has one-third the area of the electronic BZ. (c) The uniform quasihole LL with $C=+1$ folds into two degenerate zero modes at each magnetic momentum under the $V_1$ pseudopotential parent Hamiltonian. The projected screened Coulomb interaction generates dispersive quasihole bands whose connectivity depends on crystalline symmetry. A $p6$ modulation permits two isolated bands with $(C_-,C_+)=(-1,+2)$, whereas composite glide--time-reversal symmetries in a $p4gm$ symmetric system enforce band touchings, leaving a connected pair with $C_{\rm pair}=+1$.
}
\label{fig:schematic}
\end{figure}

Consider the simplest multiband case, $A_{\rm M}=A_{\rm uc}=2A_\Phi$, where the electronic LLL folds into two exactly degenerate flat bands [Fig.~\ref{fig:schematic}(a)]. At Laughlin filling $\nu_{\rm LLL}=1/3$, corresponding to $\nu_{\rm M}=2/3$ electrons per unit cell, a quasihole carries charge $e/3$ and experiences an effective flux of $2/3$ per moir\'e cell [Fig.~\ref{fig:schematic}(b)]. Consequently, three moir\'e cells enclose two quasihole flux quanta, folding the quasihole LLL into two branches in the anyon Brillouin zone [Fig.~\ref{fig:schematic}(c)]. Although degenerate under the pseudopotential parent Hamiltonian~\cite{haldane1983fractional,trugman1985exact,
ledwith2023vortexability,wang2021exact}, these branches can be split by realistic interactions, producing dispersive anyon bands.

In this Letter, we demonstrate that the resulting anyon bands exhibit the same rich topological phenomena familiar from electronic band theory. In a hexagonal moir\'e system, screened Coulomb interactions generate two isolated quasihole bands with Chern numbers $(C_-,C_+)=(-1,+2)$, distinct from the Chern number $+1$ of the parent quasihole LLL. Continuous deformations of the underlying electronic Hamiltonian can drive topological transitions through Dirac crossings, redistributing the Chern numbers among the quasihole bands even as the electronic bands remain exactly flat. Furthermore, crystalline symmetries can enforce quasihole-band crossings and connectivity, preventing the bands from being separated [Fig.~\ref{fig:schematic}(c)]. Our results establish that crystalline symmetry and band topology are organizing principles not only for electrons but also for their fractionalized excitations, opening new possibilities for anyon dynamics beyond conventional Landau-level physics.

\begin{figure}[!t]
\centering
\includegraphics[width=\linewidth]{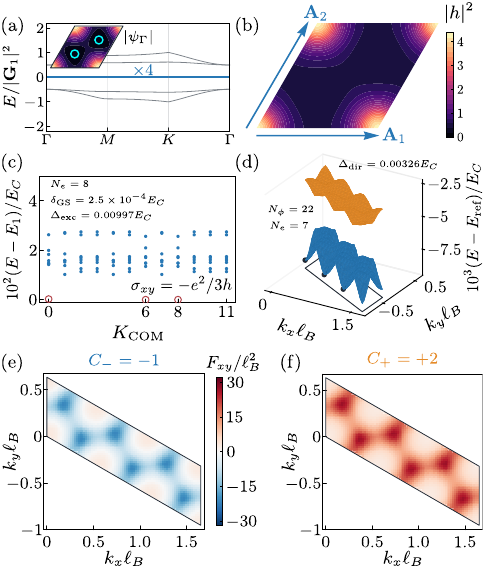}

\caption{
\textbf{Exactly flat electronic bands, Laughlin state,
and topological quasihole bands in a $p6mm$ system.} (a) Single-particle spectrum of the moir\'e-strained quadratic-band-crossing model at the original magic strain configuration, with first-shell strain coefficient $a_1\approx-2.88$ and all higher-shell coefficients zero (see End Matter). Four exactly flat bands occur at zero energy, two polarized on each sublattice. Inset: $|\psi_\Gamma(\br)|$ for one sublattice, showing the two zeros at the moir\'e-cell corners that generate the flat bands. (b) Spatial modulation $|h(\br)|^2$ relating the two sublattice-polarized bands to a reference LLL with two flux quanta per moir\'e unit cell. The primitive moir\'e lattice vectors $\bm A_1,\bm A_2$ are indicated. The magnetic length is given by $4\pi\ell_B^2 = \bm A_1\times \bm A_2$. (c) Many-body spectrum of the screened Coulomb interaction projected into the two electronic flat bands at $\nu_{\rm LLL}=1/3$ ($\nu_{\rm M}=2/3$), obtained on a symmetric 12-cell torus with $N_\phi=24$ and $N_e=8$. The three lowest states are highlighted. (d) Two quasihole bands on an 11-cell tilted torus with $N_\phi=22$ and $N_e=7$. The lower band has three magnetic-translation-related minima. (e,f) Berry curvature $F_{xy}$ of the lower and upper quasihole bands, respectively, over the magnetic anyon BZ, yielding $C_-=-1$ and $C_+=+2$. Energies are measured in units of $E_C=e^2/(\epsilon\ell_B)$. The quasihole energies are measured relative to the uniform-LLL screened-Coulomb quasihole energy $E_{\rm ref}$ for the same system size.
}
\label{fig:p6}

\end{figure}

\noindent\textit{Flat-band model and Laughlin state.}---
Although our results are generic to any multiband vortexable system, As a concrete realization of the $n=2$, $m=1$ regime, where the physical unit cell encloses two flux quanta of the reference LLL, we consider a quadratic band crossing point subject to a moir\'e-periodic strain field, described by the continuum Hamiltonian~\cite{wan2023topological,sarkar2023symmetry,sarkar2025unconventional}
\begin{equation}
\mathcal{H}(\br)=
\begin{pmatrix}
0 & \mathcal D^\dagger(\br)\\
\mathcal D(\br) & 0
\end{pmatrix},
\qquad
\mathcal D(\br)=-4\bar\partial_z^2+\widetilde A(\br),
\label{eq:single-particle}
\end{equation}
where $z=x+\ii y$ and $\widetilde A(\br) = A_x(\br)+\ii A_y(\br)$, describes the strain field with $A_x = u_{xx}-u_{yy}$ and $A_y = u_{xy}$. This Hamiltonian possesses chiral or sublattice symmetric $\{\sigma_z,\mathcal{H}(\br)\} =0$, and time reversal symmetry $\sigma_x \mathcal{K}$ (where $\mathcal{K}$ is complex conjugation) that relates the two sublattices.  As was shown in~\cite{wan2023topological}, that for $p6mm$ wallpaper group symmetric strain field, at a magic value of the strain strength, four exactly flat bands appear at zero energy, with two bands polarized on each sublattice [Fig.~\ref{fig:p6}(a)]. The two bands on each sublattice together carry Chern number $C=\pm1$. We focus on the two bands polarized on one sublattice. As shown in the inset of Fig.~\ref{fig:p6}(a), the corresponding $\Gamma$-point wavefunction $\psi_\Gamma(\br)$ has two zeros at the corners of the moir\'e unit cell. These zeros allow us to construct two linearly independent zero-energy Bloch wavefunctions by multiplying $\psi_\Gamma(\br)$ by appropriate holomorphic functions of $z$. Importantly, these two flat bands together are equivalent to a single LLL multiplied by a spatial modulation $h(\br)$. In a suitable gauge, their wavefunctions can be written as
\begin{equation}
\psi_{\bk}^{(i)}(\br)
\propto
h(\br)\psi_{\bk,i}^{\rm LLL}(\br),
\qquad i=1,2,
\label{eq:flatband-LLL}
\end{equation}
where $\psi_{\bk,i}^{\rm LLL}$ are two independent wavefunctions of an LLL with two flux quanta per moir\'e unit cell. These bands are collectively vortexable, meaning that multiplication of a wavefunction by the complex coordinate $z=x+\ii y$ preserves the combined flat-band subspace, allowing vortices to be inserted without leaving this subspace ~\cite{ledwith2023vortexability}. The modulation $|h(\br)|^2$, shown in Fig.~\ref{fig:p6}(b), encodes the spatial variation of the effective magnetic field in the Aharonov--Casher mapping, $B(\br)/B_0
=1-\ell_B^2\nabla^2\ln|h(\br)|$, where $B_0=2\Phi_0/A_{\rm M}$ is the average effective field~\cite{ledwith2020fractional}. Thus, the spatial modulation changes the effective magnetic-field distribution while preserving two flux quanta per moir\'e cell. Importantly, $|h(\br)|^2$ retains the full moir\'e periodicity and does not permit a smaller physical unit cell containing only one flux quantum. The explicit construction of $h(\br)$ and the corresponding LLL wavefunctions are given in the Supplemental Material (SM)~\cite{SM2026}.

To study the interacting states in these flat bands, we project the screened Coulomb interaction $V(\bm q) = 2\pi e^2 \tanh(d_s |\bm q|)/(\epsilon |\bm q|)$, where $d_s$ is screening length, onto the two sublattice-polarized flatbands [End Matter].~\footnote{The projection is justified when the Coulomb interaction scale is smaller than the gap between flatbands and remote bands, and when the filling fraction is $\nu_\text{M}<2$ such that exchange interaction generally drives all the electrons to one of the two sublattices.} Since each moir\'e unit cell contains two reference LLL flux quanta, the number of single-particle states (equals the number of effective flux quanta) is $N_\phi=2N_{\rm uc}$, where $N_{\rm uc}$ is the number of moir\'e unit cells. Consequently, with $N_e$ electrons, filling $\nu_{\rm M}=N_e/N_{\rm uc}=2/3$ corresponds to $\nu_{\rm LLL}=N_e/N_\phi=1/3$. As shown in Fig.~\ref{fig:p6}(c), exact diagonalization of the projected screened Coulomb interaction yields three quasi-degenerate ground states separated from excited states by a finite gap. Flux insertion through the torus gives a Hall conductivity $\sigma_{xy}=-e^2/(3h)$, consistent with a $\nu_{\rm LLL}=1/3$ Laughlin state. This state provides the starting point for investigating the quasihole excitations in the moir\'e system.

\noindent\textit{Dispersive quasihole bands.}---
We now investigate the excitation spectrum obtained by introducing a single quasihole into this Laughlin state. As illustrated in Fig.~\ref{fig:schematic}, the effective magnetic flux experienced by a quasihole is $2/3$ per moir\'e unit cell, giving rise to two quasihole states at each momentum in the magnetic aBZ. An important question is whether these two states remain degenerate. Although the underlying electronic bands are exactly flat at the magic strain strength, this fine tuning does not constrain the quasihole energies to be degenerate. For the $V_1$ pseudopotential parent Hamiltonian, all single-quasihole states are exact zero modes, even in the presence of the moir\'e modulation~\cite{ledwith2023vortexability,dong2023manybody,yan2025anyon,iyer2026dispersion}. The screened Coulomb interaction, however, generally lifts this degeneracy and produces a nontrivial quasihole dispersion.

To calculate the quasihole dispersion, we consider the one-quasihole sector with $N_\phi=3N_e+1$. Exact diagonalization (ED) of the $V_1$ pseudopotential Hamiltonian yields two zero modes $|\Psi^h_{\bk,a}\rangle$, $a=1,2$, at each quasihole momentum $\bk$ (SM~\cite{SM2026}). Projecting the screened Coulomb interaction onto this subspace and diagonalizing it gives the quasihole energy bands $E_\pm(\bk)$ and eigenstates
\begin{equation}
|\Phi_{\bk,\pm}\rangle
=
\sum_{a=1}^2
v_{\pm,a}(\bk)|\Psi^h_{\bk,a}\rangle.
\label{eq:main-anyon-eigenstates}
\end{equation}
We benchmark this zero-mode projection against full screened-Coulomb ED on $N_e=5$, $N_\phi=16$ clusters, finding high fidelities between the variational and exact low-energy quasihole subspaces, although small differences in their energies remain (see SM~\cite{SM2026}). To assign quasihole momenta to the many-electron ED states, we extend the center-of-mass (COM) momentum correspondence of Ref.~\cite{iyer2026dispersion} to general tilted torus geometries. By continuously varying the electronic boundary twists, corresponding to flux insertion through the torus, we sample the quasihole momenta throughout the magnetic aBZ and reconstruct the dispersions $E_\pm(\bk)$ (SM~\cite{SM2026}). Figure~\ref{fig:p6}(d) shows the resulting quasihole dispersion for the $p6$ modulation. Unlike the exactly degenerate electronic flat bands, the two quasihole bands acquire finite bandwidths and are separated by a direct gap.
We verified that the quasihole dispersions satisfy $E_a(\bk)=E_a(\bk+\bm G_1/3)$, as required by the quasihole magnetic translation algebra discussed in Fig.~\ref{fig:schematic}(b) (SM~\cite{SM2026}).

To characterize the quasihole-band topology, we use the localized Laughlin quasihole states $|\psi^h(\eta)\rangle$~\cite{yan2025anyon,iyer2026dispersion,SM2026} ($\eta$ is the position of the quasihole), which span the one-quasihole zero-mode space and obey the magnetic translation algebra of a charge-$e/3$ excitation. Their overlaps with the many-electron eigenstates,
\begin{equation}
\varphi_{a,\bk}(\eta)\propto
\langle\psi^h(\eta)|\Phi_{\bk,a}\rangle_e,
\label{eq:quasiholeBlochfunction}
\end{equation}
therefore describe quasihole Bloch wavefunctions. We compute the Berry curvature and Chern numbers of $\varphi_{a,\bk}(\eta)$ over the magnetic aBZ using electronic boundary twists, with the appropriate magnetic boundary conditions  (see SM~\cite{SM2026} for the details). These Chern numbers characterize the quasihole bands, rather than the electronic many-body Hall response~\cite{zeybek2022structure}. Remarkably, the Berry curvatures shown in Figs.~\ref{fig:p6}(e) and \ref{fig:p6}(f) give the Chern numbers $C_-=-1,\qquad C_+=+2,$ whose sum equals the Chern number $+1$ of the parent quasihole LLL. Thus, the moir\'e modulation does not merely generate quasihole dispersion: it splits the parent quasihole LL into isolated bands with entirely different Chern numbers. In particular, the lowest quasihole band has a Chern number opposite in sign to that of the parent LLL. The resulting dispersion and Berry curvature allow quasihole dynamics that cannot occur in a uniform LLL.

\begin{figure}[t]
\centering
\includegraphics[width=0.42\textwidth]
{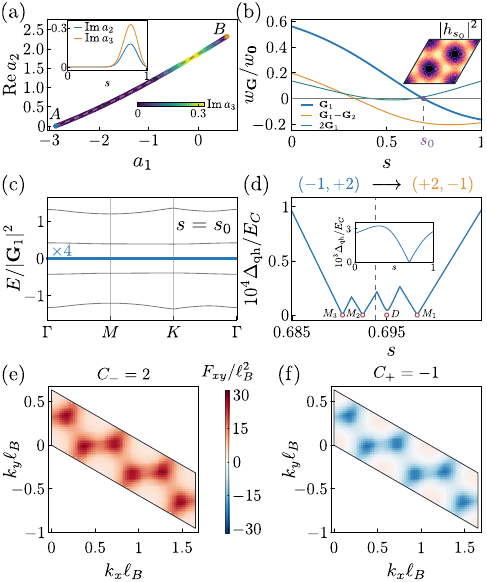}
\caption{
\textbf{Topological transitions of quasihole bands
along a continuous magic-strain path.} (a) Continuous family of $C_6$-symmetric magic strain fields connecting configurations A and B, with shell coefficients $(a_1,a_2,a_3)\approx(-2.88,0,0)$ and $(0.58,2.33,0)$, respectively. A corresponds to the strain in Fig.~\ref{fig:p6}. The projection of the path onto the $(a_1,\operatorname{Re}a_2)$ plane is shown in the main panel. The inset shows the imaginary shell coefficients along the path. The complex detour avoids the six-flat-band point of the real two-shell model. See End Matter for details. (b) Evolution of representative normalized Fourier coefficients $w_{\bG}(s)/w_{\bm0}(s)$ of $|h_s(\br)|^2$, where $\mathbf G_i\cdot\bm A_j=2\pi\delta_{ij}$. The first-shell coefficient vanishes at $s_0\simeq0.6938$, while higher harmonics remain nonzero. Inset: $|h_{s_0}(\br)|^2$. (c) Single-particle spectrum at $s=s_0$, showing four exactly flat electronic bands separated from the remote bands. (d) Minimum direct quasihole-band gap $\Delta_{\rm qh}=\min_{\bk}
[E_+(\bk)-E_-(\bk)]$ along the strain path. The main panel magnifies the transition region, while the inset shows the full range $0\leq s\leq1$. The blue curve gives the sampled momentum-mesh minimum. Red open circles mark the refined gap closings: $M_1$, $M_2$, and $M_3$ denote three inversion-invariant momentum families, each containing three magnetic-translation-related Dirac crossings, while $D$ denotes six crossings at generic momenta related by magnetic translations and inversion. The vertical dashed line marks $s=s_0$. 
(e,f) Berry curvature of the lower and upper quasihole bands at B, yielding $C_-=+2$ and $C_+=-1$, respectively. The quasihole calculations use an 11-cell tilted torus with $N_\phi=22$ and $N_e=7$; energies are measured in units of $E_C=e^2/(\epsilon\ell_B)$.
}
\label{fig:magic-strain}
\end{figure}

\noindent\textit{Topological transitions along a magic-strain path.}---
We next show that the individual quasihole-band Chern numbers can change under continuous deformations of the strain field, while the electronic bands remain exactly flat. We construct a family of $C_6$-symmetric magic strain fields $\widetilde A_s(\br)$, $0\leq s\leq1$, using the first three reciprocal shells (see End Matter). The path connects the single-shell strain configuration of Fig.~\ref{fig:p6} to a two-shell configuration, with complex higher-shell coefficients allowing us to avoid a point with six exactly flat bands and a reduced physical unit cell [Fig.~\ref{fig:magic-strain}(a)]. Expanding the resulting modulation as $|h_s(\br)|^2=\sum_{\bm G}w_{\bm G}(s)
e^{i\bm G\cdot\br}$, we find that the first-shell coefficient $w_{\bm G_1}(s)$ changes sign, vanishing at $s_0\simeq0.6938$, while higher harmonics remain nonzero [Fig.~\ref{fig:magic-strain}(b)]. Thus, the uniform LLL is never restored along the path. Figure~\ref{fig:magic-strain}(c) shows the four exactly flat electronic bands at $s=s_0$, separated from the remote bands.

Despite the persistence of exactly flat electronic bands, the quasihole bands undergo topological transitions. Their direct gap closes and reopens at four closely spaced values of $s$ near $0.69$ [Fig.~\ref{fig:magic-strain}(d)]. On the 11-cell torus, we identify fifteen Dirac crossings: nine belonging to three inversion-invariant momentum families $M_1$, $M_2$, and $M_3$, and six at generic momenta, denoted by $D$. The quasihole magnetic translation algebra relates the crossings within each threefold family, while inversion relates the two families of generic crossings. Each of the nine $M_i$ crossings transfers $+1$ to the lower-band Chern number, whereas each of the six $D$ crossings transfers $-1$, giving $\Delta C_-=9(+1)+6(-1)=+3$. At the final strain configuration, the isolated quasihole bands carry $(C_-,C_+)=(+2,-1)$ [Figs.~\ref{fig:magic-strain}(e,f)]. The agreement with the Chern-number transfer inferred from the Dirac crossings provides an independent consistency check of our method for computing Chern numbers from the quasihole Bloch wavefunctions in Eq.~\eqref{eq:quasiholeBlochfunction}. Thus, a continuous family of exactly flat electronic Hamiltonians can support distinct topological quasihole-band structures. Note that no finite $N_{\rm uc}$-torus with $2N_{\rm uc} = 3N_e+1$ satisfies $C_6$ symmetry, so the separate transition parameters and narrow intermediate Chern-number plateaus are finite-geometry results (see End Matter).

\begin{figure}[t]
\centering
\includegraphics[width=0.4\textwidth]
{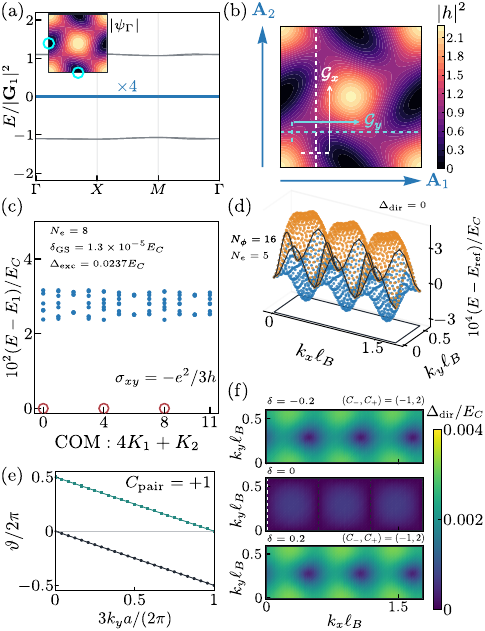}

\caption{
\textbf{Symmetry-protected quasihole-band connectivity
in a $p4gm$ symmetric system.} (a) Single-particle spectrum of the quadratic-band-crossing model with the $p4gm$-symmetric strain field $\widetilde A(\br)=
\ii\alpha[\cos(\bm G_1\cdot\br)
-\cos(\bm G_2\cdot\br)]$, where $\bm G_{1,2}=(2\pi/a)\hat{\bm x},(2\pi/a)\hat{\bm y}$ and $\alpha/|\bm G_1|^2\simeq2.24$ ~\cite{sarkar2023symmetry}. Four exactly flat bands appear at zero energy, two polarized on each sublattice. Inset: $|\psi_\Gamma(\br)|$, showing the two edge-center zeros used in the flat-band construction. (b) Spatial modulation $|h(\br)|^2$ obtained from the same strain model, with the composite glide--time-reversal operations $\mathcal G_x$ and $\mathcal G_y$ indicated. The construction of $h(\br)$ is described in the End Matter and Supplemental Material. (c) Many-body screened-Coulomb ED spectrum at $\nu_{\rm LLL}=1/3$ ($\nu_{\rm M}=2/3$), on a 12-cell torus with $N_\phi=24$ and $N_e=8$. (d) Quasihole dispersion obtained by projecting the screened Coulomb interaction into the one-quasihole $V_1$ zero-mode space on an eight-cell tilted square torus with $N_\phi=16$ and $N_e=5$. The two bands exhibit symmetry-enforced degeneracies shown in black. (e) Non-Abelian Wilson-loop spectrum of the connected quasihole-band subspace, yielding $C_{\rm pair}=+1$. (f) Direct quasihole-band gap $\Delta_{\rm dir}(\bk)=E_+(\bk)-E_-(\bk)$ for $\delta=-0.2,0,+0.2$, where $\delta$ controls a glide-breaking deformation of $|h(\br)|^2$ (see End Matter). The symmetry-protected crossings at $\delta=0$ become gapped for either sign of $\delta$, yielding $(C_-,C_+)=(-1,+2)$. The quasihole momenta refer to the magnetic anyon BZ, and energies are measured in units of $E_C=e^2/(\epsilon\ell_B)$.
}
\label{fig:p4gm}

\end{figure}

\noindent\textit{Symmetry-protected quasihole-band connectivity.}---
We finally consider the same quadratic-band-crossing model with a periodic strain field introduced in Ref.~\cite{sarkar2023symmetry} such that the system is $p4gm$ symmetric. At its magic strain strength, four exactly flat bands appear, with two polarized on each sublattice. As in the hexagonal model, the two bands on one sublattice are related to a reference LLL with two flux quanta per moir\'e unit cell. The single-particle spectrum, the two zeros of $\psi_\Gamma(\br)$, and the corresponding Laughlin-state spectrum are shown in Figs.~\ref{fig:p4gm}(a, c). The spatial modulation $|h(\br)|^2$ and the composite glide--time-reversal operations are illustrated in Fig.~\ref{fig:p4gm}(b).

Unlike the $p6$ case, here composite glide--time-reversal symmetry prevents the two quasihole bands from being globally separated. Even though the two bands polarized to the same sublattice do not possess either of glide and time-reversal symmetry, but they obey the composite antiunitary glide--time-reversal symmetry. Let $\mathcal G_x$ and $\mathcal G_y$ denote the antiunitary glide--time-reversal operations, and $\mathcal G_j^{(N_e)}$ their many-electron counterparts acting on all $N_e$ electrons. Their squares are ordinary lattice translations, $(\mathcal G_y^{(N_e)})^2
=\mathsf t_{\rm COM}(\bA_1),
\qquad
(\mathcal G_x^{(N_e)})^2
=\mathsf t_{\rm COM}(\bA_2)$ where $\mathsf t_{\rm COM}(\bA_i)$ are the COM translations by primitive moir\'e lattice vectors. On the corresponding glide-invariant COM momentum boundaries, the translation eigenvalue is $-1$. The antiunitary symmetry therefore squares to $-1$ at the COM momentum boundary and enforces a twofold Kramers degeneracy (see SM~\cite{SM2026} for details). Through the COM-to-anyon momentum correspondence, these degeneracies produce the protected quasihole-band crossings in Fig.~\ref{fig:p4gm}(d). Although the two quasihole bands cannot be assigned separate Chern numbers, their combined subspace remains well defined. The non-Abelian Wilson loop in Fig.~\ref{fig:p4gm}(e) gives $C_{\rm pair}=+1$, in agreement with the total Chern number of the parent quasihole LLL.

To further demonstrate that $\mathcal G_i$ protect the degeneracy, we introduce a perturbation of the ideal-band modulation, $|h_\delta(\br)|^2\propto
|h(\br)|^2e^{\delta f(\br)}$, where $f(\br)$ is a generic moir\'e-periodic function that breaks both $\mathcal G_i$ (End Matter). The resulting bands remain exactly flat at the single-particle level, while the quasihole-band crossings become gapped for $\delta=\pm0.2$ [Fig.~\ref{fig:p4gm}(f)], yielding isolated bands with $(C_-,C_+)=(-1,+2)$. Thus crystalline symmetry can enforce connectivity between topological quasihole bands, even when interactions would otherwise allow them to separate.

\noindent\textit{Discussion.}---
We have shown that moir\'e commensurability can reorganize Laughlin quasiholes into multiple topological bands, even when the underlying electronic bands are exactly flat. The screened Coulomb interaction generates dispersive quasihole bands whose individual Chern numbers can differ from that of the parent quasihole LLL. These Chern numbers can be redistributed through Dirac crossings along a continuous magic-strain path, while crystalline symmetries can instead enforce band connectivity. Importantly, these Chern numbers characterize fractionally charged excitations, distinct from the electronic many-body Hall response, and their Berry curvature governs the effective quasihole dynamics under slowly varying external potentials (see SM~\cite{SM2026} for details). Our results thus extend the principles of Bloch-band topology to fractionalized excitations and provide a starting point for exploring their transport and collective behavior upon doping FCIs.

\noindent\textit{Note added.}---
During completion of this work, we became aware of a related study on multiple anyon bands~\cite{xiao2026anyon}.

\let\oldaddcontentsline\addcontentsline
\renewcommand{\addcontentsline}[3]{}
\begin{acknowledgments}
\noindent \textit{Acknowledgments}.---The authors are indebted to Kai Sun for numerous discussions on related topics as well as for carefully reading the manuscript.
AI usage: OpenAI ChatGPT 5.6, 6.0 and 6.1 were used to improve the arguments as well as in finding the magic strain path. All text is human written. S.S. acknowledges support from the Laboratory for Physical Sciences through the Condensed Matter Theory Center at the University of Maryland during part of this work. X.W. acknowledges support from the UW Thouless Fellowship in Physics.

\end{acknowledgments}

\bibliographystyle{apsrev4-1}
\bibliography{ref}

\setcounter{equation}{0} 
\setcounter{figure}{0}
\renewcommand{\theequation}{A\arabic{equation}}
\renewcommand{\theHequation}{A.\arabic{equation}}
\renewcommand{\thefigure}{A\arabic{figure}}

\section*{End Matter}
\label{app:methods}

\textit{Details of the strain field used in Figs.~\ref{fig:p6} and~\ref{fig:magic-strain}.}--- For the continuous magic-strain path in Fig.~\ref{fig:magic-strain} starting from the magic strain in Fig.~\ref{fig:p6}, we include the first three reciprocal shells. We use dimensionless coordinates with the first-shell reciprocal magnitude set to unity, and choose
\begin{equation}
\mathbf G_1=\left(\frac{\sqrt3}{2},-\frac12\right),
\qquad
\mathbf G_2=(0,1).
\end{equation}
In this unit $\ell_B=(2\pi/\sqrt{3})^{1/2}$. with $u=\mathbf G_1\cdot\br$ and $v=\mathbf G_2\cdot\br$. The strain field is
\begin{equation}
\widetilde A_s(\br)
=
\frac12\sum_{j=1}^{3}a_j(s)f_j(u,v),
\label{eq:magic-strain-field}
\end{equation}
where
\begin{align}
f_1={}&
e^{-\ii\pi/3}\cos u
+e^{\ii\pi/3}\cos(u+v)-\cos v,
\nonumber\\
f_2={}&
e^{-2\ii\pi/3}\cos(u-v)
+\cos(2u+v)
\nonumber\\
&+e^{2\ii\pi/3}\cos(u+2v),
\nonumber\\
f_3={}&
e^{-\ii\pi/3}\cos(2u)
+e^{\ii\pi/3}\cos(2u+2v)
-\cos(2v).
\end{align}
These terms contain reciprocal vectors of lengths $1$, $\sqrt3$, and $2$, respectively, and obey the angular momentum-two transformation law under $C_6$ appropriate to the quadratic-band-crossing model.

We interpolate between the original strain configuration in Fig.~\ref{fig:p6} and the new magic strain configurations by choosing
\begin{align}
a_1(s)&=(1-s)a_A+sa_B,
\nonumber\\
a_3(s)&=
0.3\ii\,\frac{s(1-s)}{s_c(1-s_c)}
\exp\!\left[-\frac{(s-s_c)^2}{0.14^2}\right],
\label{eq:app-detour}
\end{align}
where
\begin{align}
a_A&=-2.87759914,\quad
a_B=0.58143052,\nonumber\\
s_c&=-\frac{a_A}{a_B-a_A}=0.83190936.
\end{align}

The remaining complex coefficient $a_2(s)$ is determined by continuously following the magic branch, starting from $a_2(0)=0$. For each $s$ and trial $a_2$, we solve
\begin{equation}
[-4\bar\partial_z^{\,2}
+\widetilde A_s(\br)]
\psi_{\Gamma,s}(\br)=0
\end{equation}
for the symmetry-protected $\Gamma$-point zero mode, and tune $a_2$ to satisfy
\begin{equation}
\psi_{\Gamma,s}
\left(\frac{\mathbf A_1+\mathbf A_2}{3}\right)=0.
\end{equation}
Inversion fixes the second zero at $2(\mathbf A_1+\mathbf A_2)/3$, allowing the construction of two exactly flat bands per sublattice. Following the Birman--Schwinger method of Ref.~\cite{becker2022mathematics, becker2021spectral,sarkar2025ideal}, we independently verify the magic condition by checking that the operator
\begin{equation}
T_{2,s}(p)=
-(2\ii\bar\partial_z-p)^{-2}
\widetilde A_s(\br)
\end{equation}
has two eigenvalues equal to unity at generic electronic momentum $\bm p$. At $s=1$, we obtain $a_2(1)=4a_B$.

The imaginary strain coefficients allow the path to avoid the real two-shell configuration where the physical unit cell becomes three times smaller and six flat bands appear. The resulting strain remains compatible with a real periodic displacement field. Extracting $h_s(\br)$ from the electronic zero modes, we find that the first harmonic of $|h_s(\br)|^2$ vanishes at $s_0=0.69379743$, while higher harmonics remain nonzero. Thus neither the uniform LLL nor a smaller physical unit cell is recovered when the first-harmonic is zero.

\textit{Details of the strain field used in Fig.~\ref{fig:p4gm}.}--- For the square-lattice $p4gm$ model in Fig.~\ref{fig:p4gm}, we use the periodic strain field ~\cite{sarkar2023symmetry}
\begin{equation}
\widetilde A(\br)
=
\ii\alpha
\left[
\cos(\bm G_1\cdot\br)
-
\cos(\bm G_2\cdot\br)
\right],
\label{eq:p4gm-strain}
\end{equation}
where $\bm G_1=(2\pi/a)(1,0)$ and $\bm G_2=(2\pi/a)(0,1)$. With this strain field the Hamiltonian in Eq.~\eqref{eq:single-particle} possesses glide and $C_4$ symmetries~\cite{sarkar2023symmetry}. At the magic value $\alpha/|\bm G_1|^2\simeq2.24$, the quadratic-band-crossing Hamiltonian has four exactly flat bands, two polarized on each sublattice. The $\Gamma$-point wavefunction has two zeros at symmetry-related edge centers of the moir\'e unit cell, which generate the two holomorphic flat-band wavefunctions.

\textit{Glide-breaking deformation of the
$p4gm$ ideal bands}-- To examine the stability of the symmetry-protected quasihole-band crossings, we deform the spatial modulation of the $p4gm$ ideal bands according to
\begin{equation}
W_\delta(\br)
=
\frac{
|h(\br)|^2e^{\delta f(\br)}
}{
\left\langle |h|^2e^{\delta f}\right\rangle
},
\label{eq:p4gm-deformation}
\end{equation}
where
\begin{equation}
\begin{split}
f(\br)={}&
\cos\theta_1+0.73\sin\theta_2\\
&+0.41\sin(\theta_1+\theta_2)
+0.27\cos(\theta_1-\theta_2),
\end{split}
\end{equation}
with $\theta_j=\bG_j\cdot\br$. Here $\langle\cdots\rangle$ denotes the spatial average over a moir\'e unit cell.

Numerically, we evaluate the Fourier coefficients
\begin{equation}
w_{\bG}^{(\delta)}
=
\left\langle
W_\delta(\br)e^{-\ii\bG\cdot\br}
\right\rangle
\end{equation}
on a $1024\times1024$ uniform real-space grid, retaining harmonics $\bG=n_1\bG_1+n_2\bG_2$ with $|n_1|,|n_2|\leq8$. The resulting truncated Fourier series defines $|h_\delta(\br)|^2$ used in the calculations.

We consider $\delta=0,\pm0.2$. The deformation breaks both glide symmetries as well as $C_2$ and $C_4$, while preserving moir\'e translation symmetry. The electronic wavefunctions are reconstructed as $\psi_{\bp,\ell}^{(\delta)}(\br)
\propto h_\delta(\br)\psi_{\bp,\ell}^{\rm LLL}(\br)$,
and their overlap matrices, interaction form factors, and quasihole zero modes are recalculated.

The bare screened Coulomb interaction and all screening parameters are held fixed. Thus the change in the quasihole dispersion originates entirely from the deformation of the electronic wavefunctions. No corresponding continuum strain field is assumed for $\delta\neq0$.

\textit{Projected Interactions.}---The interaction projected to the two sublattice-polarized flat bands is give by
\begin{equation}
\begin{split}
H_\text{int} = \frac{1}{2A} \sum_{\mathbf{k}_1\mathbf{k}_2\mathbf{k}_3\mathbf{k}_4}^\text{BZ}& c^\dagger_{\mathbf{k}_1,i}c^\dagger_{\mathbf{k}_3,j}c_{\mathbf{k}_4,k}c_{\mathbf{k}_2,l}\times\\
&\sum_\mathbf{q}V(\mathbf{q})\,[\lambda_\mathbf{q}(\mathbf{k}_1,\mathbf{k}_2)]_{il}[\lambda_{-\mathbf{q}}(\mathbf{k}_3,\mathbf{k}_4)]_{jk}
\end{split}
\end{equation}
where $A=N_{\rm uc} A_\text{M}$ is the system area containing $N_{\rm uc}$ moir\'e unit cells, $V(\mathbf{q})$ is interaction potential, and $[\lambda_\mathbf{q}(\mathbf{k},\mathbf{k}')]_{ij} = \langle\tilde{\psi}_\mathbf{k}^{(i)}|e^{-i\mathbf{q}\cdot\mathbf{r}}|\tilde{\psi}_{\mathbf{k}'}^{(j)}\rangle$ is the form factor, $\tilde{\psi}_\mathbf{k}^{(i)}$ are the orthonormalized versions of the single particle wavefunctions given in Eq.~\eqref{eq:flatband-LLL}. The screened interaction is
\begin{equation}
V_{\rm sc}(q)=
\frac{2\pi e^2}{\epsilon q}
\tanh(qd_{\rm sc}),
\label{eq:app-screening}
\end{equation}
with the finite $q=0$ limit understood. Throughout we use screening length $d_{\rm sc}=2/|\bm G_1|$, where $\bm G_1$ is a shortest moir\'e reciprocal lattice vector. On the other hand, the $V_1$ pseudopotential is given by $V(\mathbf{q})\propto -q^2$.

\textit{Finite-size geometries.}--- For the ED many-body spectra at $\nu_{\rm LLL}=1/3$ ($\nu_{\rm M}=2/3$), we use 12-cell tori with $N_\phi=24$ and $N_e=8$: a $C_6$-symmetric torus for $p6$ and a $3\times4$ moir\'e-cell torus for $p4gm$. For the ED one-quasihole calculations, $N_\phi=3N_e+1$, we use an 11-cell tilted torus with $\bm L_1=3\bm A_1+\bm A_2$ and $\bm L_2=-2\bm A_1+3\bm A_2$ for $p6$ ($N_\phi=22$, $N_e=7$), and an eight-cell tilted square torus with $\bm L_1=2\bm A_1+2\bm A_2$ and $\bm L_2=-2\bm A_1+2\bm A_2$ for $p4gm$ ($N_\phi=16$, $N_e=5$). The 11-cell torus preserves $C_2$ but not the full $C_6$ symmetry. In fact, no $C_6$-symmetric finite torus is compatible with the one-quasihole sector. This is because: for the triangular lattice, a $C_6$-symmetric torus requires $N_{\rm uc}=p^2+pq+q^2$, with integers $p,q$, implying $N_{\rm uc}=0$ or $1\pmod 3$. However, the one-quasihole condition $2N_{\rm uc}=3N_e+1$ requires $N_{\rm uc}=2\pmod 3$.

\textit{Dirac crossings and Chern-number transfer.}--- Along the magic-strain path, the two quasihole bands change their Chern numbers from $(-1,+2)$ to $(+2,-1)$. The quasihole magnetic translation algebra implies the energy-periodicity relation $E_\pm(\bk)=E_\pm(\bk+\bm G_1/3)$. Consequently, an isolated band touching has two additional magnetic-translation partners.

The lower band also reverses its inversion character between the endpoints at three inequivalent inversion-invariant momentum families. These changes require at least nine gap-closing momenta. For the strain path of Fig.~3, the numerical calculation finds fifteen Dirac points: nine in the three inversion-invariant families, and six additional points related by magnetic translations and inversion. The critical strain parameters are $s=0.690351$, $0.692477$, $0.695009$, and $0.698222$, respectively.

The individual Dirac charges are checked both by the linearized two-band Hamiltonian and by integrating the quasihole Berry flux around small closed contours in $(k_x,k_y)$ space. The nine inversion-invariant-momentum crossings carry charge $+1$ each, and the six additional crossings carry $-1$ each, giving
\begin{equation}
\Delta C_-=9-6=3,\qquad
\Delta C_+=-3.
\end{equation}
The total Chern number of the two-band subspace remains $+1$.

The strain preserves $C_6$, but the 11-cell torus preserves only $C_2$. Thus the three inversion-invariant momentum families need not close at the same parameter on this finite torus, although they are rotation-related in the $C_6$-symmetric bulk construction. The narrow intermediate Chern-number plateaus should not be interpreted as established bulk phases.

\clearpage
\let\addcontentsline\oldaddcontentsline
\onecolumngrid

\counterwithout{equation}{section}
\setcounter{secnumdepth}{2}
\makeatletter
\renewcommand \thesection{S-\@arabic\c@section}
\renewcommand\thetable{S\@arabic\c@table}
\renewcommand \thefigure{S\@arabic\c@figure}
\renewcommand \theequation{S\@arabic\c@equation}
\renewcommand{\theHequation}{S.\arabic{equation}}
\makeatother
\setcounter{equation}{0}  
\setcounter{figure}{0}  
\setcounter{section}{0}  
\counterwithin{figure}{section} 
\begin{center}
    \bfseries\large Supplemental Material
\end{center}
\tableofcontents

\section{Introduction}
This Supplemental Material provides the theoretical construction and numerical details underlying the quasihole-band results presented in the main text. We first construct the Laughlin quasihole zero modes on a torus and derive their magnetic translation properties, including the correspondence between electronic center-of-mass and quasihole momenta for general tilted geometries. We then describe the construction of two-flux ideal Chern bands, the projection of screened Coulomb interactions onto the $V_1$ pseudopotential zero-mode space, and comparisons with full exact diagonalization. Finally, we establish the symmetry constraints on quasihole-band connectivity and derive the Berry curvature and Chern numbers from quasihole Bloch wavefunctions, distinguishing them from electronic boundary-twist Chern numbers. Throughout, we consider the Laughlin state at $\nu_{\rm LLL}=1/3$, retaining the general denominator $m$ where appropriate.

\section{Laughlin quasihole zero modes on a torus}
\label{sec:zero-modes}

We consider $\Ne$ electrons and one quasihole at filling $\nu=1/m$, with
\begin{equation}
\Nphi=m\Ne+1,\qquad
\operatorname{Im}(\overline L_1L_2)
=2\pi\Nphi\ell_B^2,\qquad
\tau=\frac{L_2}{L_1},\quad \operatorname{Im}\tau>0.
\label{eq:particle-flux}
\end{equation}
Here $L_1,L_2$ are the complex torus periods. We choose coordinates such that $L_1=L_x$ is real and positive, while $L_2=L_{2x}+\ii L_{2y}$ may have a nonzero real part. We write $z_i=x_i+\ii y_i$, $\eta=\eta_x+\ii\eta_y$, and $Z=\sum_i z_i$. The Jacobi theta function satisfies
\begin{equation}
\vartheta_1(w+1\vert\tau)
=-\vartheta_1(w\vert\tau),\qquad
\vartheta_1(w+\tau\vert\tau)
=-e^{-\ii\pi\tau-2\pi\ii w}
\vartheta_1(w\vert\tau).
\label{eq:theta-identities}
\end{equation}

For an ideal Chern band, the single-particle wavefunctions can be written as
\begin{equation}
\psi_{\bk}^h(\bm r)
=\mathcal N_{\bk}\psi_{\bk}^{\rm LLL}(\bm r)h(\bm r),
\qquad
h(\bm r)
=e^{-\mathcal K(\bm r)/(4\ell_B^2)}\chi(\bm r),
\label{eq:ideal-spinor-wavefunction}
\end{equation}
where $\mathcal K(\bm r)$ is a periodic K\"ahler potential and $\chi(\bm r)$ is a normalized internal spinor with $n_c$ components,
\begin{equation}
\chi(\bm r)=
\begin{pmatrix}
\chi_1(\bm r)\\
\vdots\\
\chi_{n_c}(\bm r)
\end{pmatrix},
\qquad
\chi^\dagger(\bm r)\chi(\bm r)=1.
\label{eq:chi-normalization}
\end{equation}
Consequently, $h^\dagger(\bm r)h(\bm r)
=e^{-\mathcal K(\bm r)/(2\ell_B^2)}$. We also define
\begin{equation}
g(\eta)
=e^{-\mathcal K(\eta)/(4m\ell_B^2)}.
\label{eq:g-factor}
\end{equation}

Following Ref.~\cite{iyer2026dispersion}, the localized Laughlin quasihole wavefunction in the analytic topological sector $\sigma=1,\ldots,m$ is
\begin{align}
\psi_\sigma^h(\{z_i\},\eta)
={}& A_\sigma
\exp\!\left[
\frac{\mathcal B(\eta)}{4m\ell_B^2}
+\sum_{i=1}^{\Ne}
\frac{\mathcal B(z_i)}{4\ell_B^2}
\right]
e^{\ii K_\sigma(Z+\eta/m)}
\prod_{\gamma=1}^{m}
\vartheta_1\!\left(
\frac{Z+\eta/m-Z^\gamma_\sigma}{L_1}
\Big\vert\tau\right)
\nonumber\\
&\times
\prod_{i=1}^{\Ne}
\vartheta_1\!\left(
\frac{z_i-\eta}{L_1}\Big\vert\tau\right)
\prod_{i<j}
\vartheta_1\!\left(
\frac{z_i-z_j}{L_1}\Big\vert\tau\right)^m
\prod_{i=1}^{\Ne}\chi(\bm r_i),
\label{eq:qh-torus}\\
\mathcal B(z)
={}&z^2-|z|^2-\mathcal K(\bm r),\qquad
K_\sigma=K_1+\frac{2\pi(\sigma-1)}{L_1},
\qquad
Z^\gamma_\sigma
=\frac{Z_1+\gamma-1}{m}L_1
-\frac{\sigma-1}{m}L_2.
\label{eq:qh-parameters}
\end{align}
The product over $\chi(\bm r_i)$ denotes a tensor product of the internal spinors of the electrons. Denoting by $\psi_\sigma^0(\{z_i\},\eta)$ the uniform scalar LLL quasihole wavefunction obtained from Eq.~\eqref{eq:qh-torus} by setting $\mathcal K=0$ and $\chi=1$, we can equivalently write
\begin{equation}
\boxed{
\psi_\sigma^h(\{z_i\},\eta)
=
g(\eta)
\left[\prod_{i=1}^{\Ne}h(\bm r_i)\right]
\psi_\sigma^0(\{z_i\},\eta).
}
\label{eq:ideal-quasihole-map}
\end{equation}

In an FCI without an external magnetic field, the spinor $\chi(\bm r)$ cancels the magnetic quasiperiodicity of the auxiliary LLL wavefunction. For primitive vectors $\bm a_1,\bm a_2$ satisfying $\bm a_1\times\bm a_2=2\pi\ell_B^2$, we choose
\begin{equation}
\chi(\bm r+\bm a_j)
=
e^{-\ii\bm a_j\times\bm r/(2\ell_B^2)}
\chi(\bm r),
\label{eq:chi-quasiperiodicity}
\end{equation}
where constant translation phases are absorbed into the Bloch-momentum convention. Since the scalar LLL wavefunction transforms with the opposite coordinate-dependent phase,
\begin{equation}
\begin{split}
&\psi_{\bk}^{\rm LLL}(\bm r+\bm a_j)
=e^{\ii\bk\cdot\bm a_j}
e^{\ii\bm a_j\times\bm r/(2\ell_B^2)}
\psi_{\bk}^{\rm LLL}(\bm r),\\
&\chi(\bm r+\bm a_j)
\psi_{\bk}^{\rm LLL}(\bm r+\bm a_j)
=e^{\ii\bk\cdot\bm a_j}
\chi(\bm r)\psi_{\bk}^{\rm LLL}(\bm r).
\end{split}
\label{eq:electron-phase-cancellation}
\end{equation}
Thus the electronic wavefunctions have ordinary Bloch translation properties whenever $\mathcal K(\bm r)$ is invariant under the corresponding lattice translation. The quasihole coordinate $\eta$ has no internal spinor factor, and its magnetic translation algebra remains projective.

The factor $A_\sigma$ fixes the relative factors between the analytic sectors. We use
\begin{equation}
A_\sigma=
\exp\!\left[
\frac{\ii K_1L_2}{m}\sigma
+\frac{\ii\pi\tau}{m}\sigma(\sigma-2)
\right],
\label{eq:A-sigma}
\end{equation}
following Ref.~\cite{iyer2026dispersion}. The parameters $K_1$ and $Z_1$ are fixed by the electron boundary twists $\theta_j$,
\begin{equation}
e^{\ii K_1L_1}
=(-1)^{\Nphi}e^{\ii\theta_1},\qquad
e^{\ii K_1L_2+2\pi\ii Z_1}
=(-1)^{\Nphi+m-1}e^{\ii\theta_2}.
\label{eq:electron-bc}
\end{equation}
These relations follow by translating one electron through either torus period and applying Eq.~\eqref{eq:theta-identities}, with the magnetic phases of the scalar LLL and internal spinor cancelling each other. The boundary conditions must be retained when the twists are varied.

For a fixed electron coordinate $z_i$, the analytic part of Eq.~\eqref{eq:qh-torus} has $m(\Ne-1)$ zeros from the electron--electron factors, one zero at $z_i=\eta$, and $m$ zeros from the COM theta function. Their total is
\begin{equation}
m(\Ne-1)+1+m=\Nphi.
\end{equation}
The remaining factors $h(\bm r_i)$ and $g(\eta)$ do not remove these zeros. For $m=3$, the wavefunctions are exact zero modes of the corresponding local $V_1$ pseudo-potential parent Hamiltonian~\cite{trugman1985exact,haldane1985periodic},
\begin{equation}
H_{V_1}\ket{\psi_\sigma^h(\eta)}=0,
\qquad
\dim\mathcal Z_0=\Nphi.
\label{eq:parent-space}
\end{equation}
For larger odd $m$, the parent interaction excludes all odd relative-angular-momentum channels below $m$, and the one-quasihole zero-mode space again has dimension $\Nphi$.

The $m$ choices of $\sigma$ do not give $m\Nphi$ independent states. Translating the quasihole around a torus period changes its analytic topological sector. As shown below, fixing one sector and allowing $\eta$ to range over an enlarged quasihole-coordinate torus spans the complete $\Nphi$-dimensional zero-mode space.

Finally, Eq.~\eqref{eq:ideal-quasihole-map} makes clear that multiplication by $h(\bm r_i)$ maps the uniform quasihole zero modes into those of the corresponding local ideal-band parent Hamiltonian. This statement does not apply to the eigenstates of an arbitrary screened Coulomb interaction. The factor $g(\eta)$ is independent of the electron coordinates and hence does not change the electronic zero-mode subspace at fixed $\eta$. However, $|g(\eta)|^2$ enters the quasihole-coordinate overlaps and changes the local Berry curvature, as discussed in Sec.~\ref{sec:chern}.

\subsection{Quasihole magnetic translations, quasihole Bloch states and quasihole Brillouin zone for aligned torus with one flux quantum per unit cell }
\label{subsec:rectangular}

We first review the construction of quasihole Bloch states for a torus whose periods are aligned with the primitive vectors of the one-flux lattice, following Ref.~\cite{iyer2026dispersion}. The quasihole carries charge $1/m$ of the electron charge magnitude. In the symmetric-gauge convention of Eq.~\eqref{eq:qh-torus}, a translation of its coordinate acts on the quasihole wavefunction as
\begin{equation}
[T_\eta(\ba)f](\eta)
=e^{-\ii\operatorname{Im}(\eta\overline a)/(2m\ell_B^2)}
f(\eta+a),
\qquad
T_\eta(\ba)T_\eta(\bb)
=e^{\ii\ba\times\bb/(m\ell_B^2)}
T_\eta(\bb)T_\eta(\ba).
\label{eq:anyon-translation}
\end{equation}
Here $\ba\times\bb=a_xb_y-a_yb_x$. These operators act on the translated quasihole wavefunctions; they do not differentiate or translate an $\eta$-dependent normalization factor introduced after forming a Bloch combination~\cite{iyer2026dispersion}.

We take the primitive lattice vectors $\mathbf a_1$ and $\mathbf a_2$ to satisfy $\mathbf a_1\times\mathbf a_2=2\pi\ell_B^2$. In this subsection, we assume that the torus periods are integer multiples of these vectors,
\begin{equation}
\mathbf L_1=N_1\mathbf a_1,\qquad
\mathbf L_2=N_2\mathbf a_2,\qquad
N_1N_2=N_\phi.
\label{eq:rect-geometry}
\end{equation}
The primitive vectors need not be perpendicular, so this construction applies equally to rectangular and oblique tori. The essential simplification is that each torus period is parallel to one primitive lattice vector, allowing the boundary conditions to be imposed separately along the two directions.

With $T_j=T_\eta(\mathbf a_j)$, the magnetic translation algebra becomes
\begin{equation}
T_1T_2=\omega T_2T_1,\qquad
\omega=e^{2\pi\ii/m},\qquad
[T_1,T_2^m]=0.
\label{eq:T12}
\end{equation}
A quasihole Bloch state is obtained by summing the localized states over the commuting translations:
\begin{equation}
\ket{\mathcal B_{\bm\kappa}(\eta)}
=
\sum_{n_1=0}^{N_1-1}\sum_{n_2=0}^{N_2-1}
e^{-2\pi\ii(\kappa_1n_1+m\kappa_2n_2)}
T_1^{n_1}T_2^{mn_2}\ket{\psi_\sigma(\eta)}.
\label{eq:rect-bloch}
\end{equation}
After removing its scalar $\eta$-dependent factor and normalizing in the electronic many-body Hilbert space, Eq.~\eqref{eq:rect-bloch} defines a normalized many-electron state. It obeys
\begin{equation}
T_1\ket{\Psi_{\bm\kappa}}
=e^{2\pi\ii\kappa_1}\ket{\Psi_{\bm\kappa}},\qquad
T_2^m\ket{\Psi_{\bm\kappa}}
=e^{2\pi\ii m\kappa_2}\ket{\Psi_{\bm\kappa}},
\quad
0\leq\kappa_1<1,\quad
0\leq\kappa_2<\frac1m.
\label{eq:rect-bz}
\end{equation}
The corresponding magnetic anyon Brillouin zone has area $2\pi/(m\ell_B^2)$, smaller than the electron Brillouin zone by a factor of $m$. Translating by $T_2$ changes $\kappa_1$ by $1/m$, connecting the different analytic topological sectors. It does not create extra independent states at a fixed electron twist.

For the aligned rectangular torus, the momentum quantization follows directly from translations through the complete periods $\mathbf L_1=N_1\mathbf a_1$ and $m\mathbf L_2=mN_2\mathbf a_2$. Define
\begin{equation}
\alpha_\sigma=\frac{K_\sigma L_1}{m}+\pi(\Ne+1),\qquad
\beta=\theta_2+\pi(m-1).
\label{eq:ab-boundaries}
\end{equation}
Applying Eq.~\eqref{eq:theta-identities} to Eq.~\eqref{eq:qh-torus} yields
\begin{equation}
T_\eta(\mathbf L_1)\ket{\psi_\sigma}
=e^{\ii\alpha_\sigma}\ket{\psi_\sigma},\qquad
T_\eta(m\mathbf L_2)\ket{\psi_\sigma}
=e^{\ii\beta}\ket{\psi_\sigma}.
\label{eq:cover-bc}
\end{equation}
A single $\mathbf L_2$ translation changes $\sigma$, while $m\mathbf L_2$ returns to the same analytic sector. Since $T_\eta(\mathbf L_1)=T_1^{N_1}$ and $T_\eta(m\mathbf L_2)=T_2^{mN_2}$ in this geometry, the allowed momenta satisfy
\begin{equation}
e^{2\pi\ii N_1\kappa_1}=e^{\ii\alpha_\sigma},
\qquad
e^{2\pi\ii mN_2\kappa_2}=e^{\ii\beta}.
\end{equation}
Thus,
\begin{equation}
\kappa_1=\frac{n_1}{N_1}
+\frac{\alpha_\sigma}{2\pi N_1},\qquad
\kappa_2=\frac{n_2}{mN_2}
+\frac{\beta}{2\pi mN_2},
\label{eq:rect-grid}
\end{equation}
with $0\leq n_1<N_1$ and $0\leq n_2<N_2$. These are the rectangular-torus results of Ref.~\cite{iyer2026dispersion}.

These results follow from the aligned geometry $\mathbf L_i=N_i\mathbf a_i$ and hold for arbitrary angles between the primitive lattice vectors. For a general tilted cluster, the torus periods are instead
\begin{equation}
\mathbf L_i=\sum_{j=1}^2 M_{ij}\mathbf a_j,
\qquad \det M=N_\phi,
\end{equation}
where $M$ is an integer matrix that need not be diagonal. Translations through the torus periods then involve both $T_1$ and $T_2$, together with the phases arising from their noncommutativity. Consequently, the momentum quantization in Eq.~\eqref{eq:rect-grid} must be generalized. We derive this generalization in the following subsection.

\section{Magnetic anyon Bloch states on a tilted torus with one flux quantum per unit cell}
\label{sec:tilted}

\subsection{Torus matrix, topological sectors, and the covering coordinate}

Take primitive vectors $\bm a_1,\bm a_2$ of a one-flux reference lattice, with $\bm a_1\times\bm a_2=2\pi\ell_B^2$. The electron torus is specified by an integer matrix $M$ with row vectors
\begin{equation}
 \begin{pmatrix}\bm L_1\\\bm L_2\end{pmatrix}
 =M\begin{pmatrix}\bm a_1\\\bm a_2\end{pmatrix},\qquad
 \det M=\Nphi=m\Ne+1.
 \label{eq:M-definition}
\end{equation}
A generic $M$ has off-diagonal entries, even after rotating $\bm L_1$ onto the $x$ axis. The torus periods, and not merely their projections on two Cartesian axes, determine the magnetic translations and the boundary twists.

The fixed-$\sigma$ quasihole kernel has periods $\bm L_1$ and $m\bm L_2$ by Eq.~\eqref{eq:cover-bc}. It can therefore be represented as a scalar LLL on a \emph{covering torus} of periods
\begin{equation}
 \bm C_1=\bm L_1,\qquad \bm C_2=m\bm L_2,\qquad
 \qell^2=m\ell_B^2,
 \quad \bm C_1\times\bm C_2=2\pi\qell^2\Nphi.
 \label{eq:cover-definition}
\end{equation}
The cover contains precisely $\Nphi$ scalar LLL orbitals. Successive strips of the cover represent the changes of $\sigma$ under $\eta\to\eta+L_2$; the cover is not an enlarged electron system. A coefficient function on this cover is an explicit representation of a physical state in the $\Nphi$-dimensional parent space.

There is a second equivalent way to retain all sectors that is particularly useful for calculating translation characters on arbitrary $M$. In integer primitive coordinates the cover periods are the rows of
\begin{equation}
 P=\begin{pmatrix}M_{11}&M_{12}\\mM_{21}&mM_{22}\end{pmatrix}.
 \label{eq:P-definition}
\end{equation}
Choose representatives of $\mathbb Z^2/P^T\mathbb Z^2$. There are $m\Nphi$ such representatives. For an ordered translation $\mathcal T(\bm r)=T_1^{r_1}T_2^{r_2}$, the product rule is
\begin{equation}
 \mathcal T(\bm r)\mathcal T(\bm s)
 =e^{-2\pi\ii r_2s_1/m}\mathcal T(\bm r+\bm s).
 \label{eq:ordered-law}
\end{equation}
Starting from one analytic sector, one can evaluate the unwrapped states $\mathcal T(\bm r)\ket{\psi_{\sigma=1}(\eta)}$ for all these representatives. Reduction modulo $P$ uses the theta-function boundary relations and their cocycle phases. Since a shift by $\bm L_2$ changes the topological sector, the $m\Nphi$ representatives are redundant descriptions of the same $\Nphi$-dimensional physical quasihole space, not $m\Nphi$ orthogonal electron states. In particular, assigning $\sigma$ from one component of $\bm r$ without simultaneously transforming the coordinate and magnetic phases does not give the same representation.

\subsection{Commuting translations and allowed tilted momenta}
\label{subsec:tilted-momenta}

Let $B$ be an integer matrix whose rows are the primitive-coordinate vectors of a quasihole magnetic cell, such that
\begin{equation}
 P=HB,\qquad H\in\mathrm{Mat}_{2\times2}(\mathbb Z).
 \label{eq:PBH}
\end{equation}
For systems with one flux quantum per unit cell, take $\det B=m$, so $\det H=\Nphi$. The translations through the rows $\bm a_1^\mathrm{qh},\bm a_2^\mathrm{qh}$ of $B$ commute because their area encloses one flux quantum for the quasihole. We define their eigenvalues by
\begin{equation}
 T_\eta(\bm a_j^\mathrm{qh})\ket{\Psi_{\bm u}}=e^{2\pi\ii u_j}\ket{\Psi_{\bm u}},\qquad
 \bk=2\pi (B a)^{-1}\bm u,
 \label{eq:B-eigen}
\end{equation}
where $a$ is the matrix with rows $\bm a_1,\bm a_2$, and vectors are treated as columns when multiplied by matrices. Here and below a formula involving reciprocal matrices assumes the same row convention on both sides. The determinant of the reciprocal magnetic cell gives the physical magnetic-BZ area $2\pi/(m\ell_B^2)$. We further define the reciprocal vectors $\bm g_1^\mathrm{qh},\bm g_2^\mathrm{qh}$ as the quasihole reciprocal vectors, satisfying $\bm g_i^\mathrm{qh}\cdot\bm a_j^\mathrm{qh}=2\pi\delta_{ij}$. Then any vector in the quasihole magnetic BZ can be written as $\bk= \sum_{i=1}^2u_i\bm g_i^\mathrm{qh} = \sum_{i=1}^2\kappa_i\bm g_i$ where $\bm g_i$ are the reciprocal vectors of the original unit cell spanned by $\bm a_1,\bm a_2$.

The torus boundary characters follow by expressing $\bm C_j=H_{j1}\bm a_1^\mathrm{qh}+H_{j2}\bm a_2^\mathrm{qh}$. Because translations are magnetic, the phase of the product of the two generators must be retained. 
Although $T_\eta(\bm b_1)$ and $T_\eta(\bm b_2)$ commute, their product differs from a translation by the sum of the displacement vectors. Since $\bm a_1^\mathrm{qh}\times\bm a_2^\mathrm{qh}=2\pi m\ell_B^2$, the magnetic translation composition law gives
\begin{equation}
T_\eta(p\bm a_1^\mathrm{qh}+q\bm a_2^\mathrm{qh})
=e^{-\ii\pi pq}
T_\eta(\bm a_1^\mathrm{qh})^pT_\eta(\bm a_2^\mathrm{qh})^q
\end{equation}
for integers $p,q$. Applying this relation to $\bm C_j=H_{j1}\bm a_1^\mathrm{qh}+H_{j2}\bm a_2^\mathrm{qh}$, and imposing the boundary characters in Eq.~\eqref{eq:cover-bc}, we obtain
\begin{equation}
H\bm u=
\frac{1}{2\pi}
\begin{pmatrix}\alpha_\sigma\\\beta\end{pmatrix}
+\frac12
\begin{pmatrix}
H_{11}H_{12}\\
H_{21}H_{22}
\end{pmatrix}
\pmod{\mathbb Z^2}.
\label{eq:tilted-grid}
\end{equation}
The distinct solutions are obtained by adding an integer vector inside the brackets before applying $H^{-1}$, with $\bm n$ taken modulo $H\mathbb Z^2$. Their number is $|\det H|=N_\phi$.

The sign of the half-integer cocycle is immaterial modulo one, but its presence matters when the product is odd. Equation~\eqref{eq:tilted-grid} yields $\Nphi$ distinct allowed $\bm u$ values. It reduces to Eq.~\eqref{eq:rect-grid} when the torus and magnetic cell are aligned.

Having determined the allowed momenta, we can construct the corresponding quasihole Bloch states by summing the localized quasihole wavefunctions over the commuting magnetic translations. Let $\mathcal V_j=T_\eta(\bm a_j^{\rm qh})$, and let $\mathcal R_H$ denote a set of representatives of $\mathbb Z^2/H^T\mathbb Z^2$. For each allowed $\bm u$, we define
\begin{equation}
\boxed{
\ket{\mathcal B_{\bm u}(\eta)}
=
\sum_{\bm n\in\mathcal R_H}
e^{-2\pi\ii\bm n^T\bm u}
\mathcal V_1^{n_1}\mathcal V_2^{n_2}
\ket{\psi_{\sigma=1}(\eta)}.
}
\label{eq:tilted-bloch}
\end{equation}
The sum contains $|\det H|=\Nphi$ terms. Since $\mathcal V_1$ and $\mathcal V_2$ commute, the boundary conditions in Eq.~\eqref{eq:tilted-grid} ensure that the sum is independent of the choice of representatives. Shifting the summation index then gives
\begin{equation}
\mathcal V_j\ket{\mathcal B_{\bm u}(\eta)}
=
e^{2\pi\ii u_j}
\ket{\mathcal B_{\bm u}(\eta)}.
\label{eq:tilted-bloch-eigen}
\end{equation}
For a nonvanishing Bloch sum, its dependence on $\eta$ is an overall scalar factor multiplying a many-electron state with definite quasihole momentum $\bm u$. Removing this scalar factor and normalizing gives the corresponding many-electron quasihole Bloch state $\ket{\Psi_{\bm u}}$.

For a physical cell that encloses $f$ original flux quanta, choose instead $\det B=fm$ and $\det H=\Nphi/f$ (provided $f$ divides $\Nphi$). The cocycle term in Eq.~\eqref{eq:tilted-grid} is then multiplied by $f$. In particular, for $f=2$ it is integral and can be dropped modulo one. There are $\Nphi/2$ magnetic momenta and two independent quasihole states at each momentum. This is the folded construction used below.

\subsection{Electron COM momentum versus quasihole momentum}
\label{subsec:affine-map}

In ED, many-electron states are labelled by their center-of-mass (COM) momentum, whereas the quasihole Bloch states constructed above are labelled by the eigenvalues of quasihole magnetic translations. We now establish the relation between these two momentum labels for a general tilted torus. The relation follows directly from the torus Laughlin quasihole wavefunction, in particular from the dependence of its COM theta function on $Z+\eta/m$, where $Z=\sum_i z_i$.

We define the electronic COM momentum $\widetilde{\bm\kappa}$ through the eigenvalues of the COM (magnetic) translations,
\begin{equation}
\mathcal U_e(\bm a_j)\ket{\Psi}
=e^{2\pi\ii\widetilde\kappa_j}\ket{\Psi},
\end{equation}
where $\mathcal U_e(\bm a_j)$ translates all electrons by the vector $\bm a_j$. For definiteness, we use the symmetric-gauge convention
\begin{equation}
[\mathcal U_e(\bm d)\Psi^0](\{z_i\})
=e^{-\ii\sum_i\operatorname{Im}(z_i\bar d)/(2\ell_B^2)}
\Psi^0(\{z_i+d\})
\label{eq:electron-mto}
\end{equation}
for the underlying uniform LLL quasihole wavefunction $\Psi^0$. The quasihole momentum is instead defined by the commuting translations in Eq.~\eqref{eq:B-eigen}. We express it in the reciprocal basis of the original one-flux lattice using $\bm u=B\bm\kappa$, so that
\begin{equation}
T_\eta(\bm b_j)\ket{\Psi_{\bm\kappa}}
=e^{2\pi\ii(B\bm\kappa)_j}
\ket{\Psi_{\bm\kappa}}.
\end{equation}
where $B$ was defined around Eq.~\eqref{eq:PBH}.

Consider translating every electron by $\bm a_j$. The electron COM coordinate changes as $Z\to Z+\Ne a_j$, while the quasihole factors $\vartheta_1((z_i-\eta)/L_1|\tau)$ transform as though the quasihole coordinate were shifted to $\eta-a_j$. However, the arguments of the COM theta function in these two transformations differ by
\begin{equation}
\left(Z+\Ne a_j+\frac{\eta}{m}\right)
-\left(Z+\frac{\eta-a_j}{m}\right)
=\frac{\Nphi}{m}a_j,
\end{equation}
where we used $\Nphi=m\Ne+1$. The two transformations can therefore be related by using the quasiperiodicity of the COM theta function, provided the additional displacement is an integer combination of the torus periods.

To achieve this, we further shift the quasihole coordinate by full torus periods, $\eta\to\eta-a_j-p_jL_1-q_jL_2$, choosing integers $p_j,q_j$ such that
\begin{equation}
\frac{\Nphi a_j+p_jL_1+q_jL_2}{m}
=U_{j1}L_1+U_{j2}L_2,
\label{eq:com-compensation}
\end{equation}
where $U_{j1},U_{j2}$ are also integers. Equation~\eqref{eq:com-compensation} ensures that the change in the COM theta-function argument is a complete torus-period translation, whose phase is known.

For a tilted torus, the periods are related to the primitive lattice vectors by $\bm L_i=\sum_j M_{ij}\bm a_j$, with $\det M=\Nphi$. Expressing Eq.~\eqref{eq:com-compensation} in primitive coordinates and multiplying by $M^{-1}$ gives
\begin{equation}
\Nphi(M^{-1})_{j\ell}
+(p_j,q_j)_\ell
=mU_{j\ell}.
\label{eq:com-integer}
\end{equation}
The first term is an integer because $J=\Nphi M^{-1}=\operatorname{adj}M$ is an integer matrix. We can therefore satisfy Eq.~\eqref{eq:com-integer} by taking $(p_j,q_j)$ to be the negative remainders of the $j$th row of $J$ modulo $m$. Collecting the two rows gives
\begin{equation}
J=\operatorname{adj}M,\qquad
Q=-\operatorname{mod}(J,m),\qquad
U=\frac{J+Q}{m}.
\label{eq:tilted-RU}
\end{equation}
The remainder is taken entrywise in $\{0,\ldots,m-1\}$. By construction, $U$ has integer entries.

The displacement of the quasihole needed to compensate an electron translation by $\bm a_j$ is
\begin{equation}
\bm\rho_j=\bm a_j+p_j\bm L_1+q_j\bm L_2.
\end{equation}
Writing $\rho_j$ in terms of unit cell vectors $\bm a_1, \bm a_2$, the coefficients of this expansion form the $j$th row of the integer matrix
\begin{equation}
r=I+QM.
\label{eq:com-r}
\end{equation}
Using $J+Q=mU$ and $JM=\Nphi I$, we find
\begin{equation}
r=I+(mU-J)M
=m(UM-\Ne I).
\label{eq:r-divisibility}
\end{equation}
Thus every entry of $r$ is divisible by $m$. On the other hand we can write $\bm\rho_j$ in terms of quasihole magnetic-cell vectors as $\bm\rho_j=n_{j1}\bm b_1+n_{j2}\bm b_2$. Then, since the quasihole magnetic cell has $\det B=m$, each displacement $\bm\rho_j$ is an integer combination of quasihole magnetic-cell vectors. In particular, if $\bm r_j$ denotes the $j$th row of $r$, its magnetic-cell coordinates are
\begin{equation}
\bm n_j=\bm r_j B^{-1}\in\mathbb Z^2.
\end{equation}

We can now evaluate the effect of the compensating quasihole translation on a Bloch state. Using the magnetic translation composition law and $\bm\rho_j=n_{j1}\bm b_1+n_{j2}\bm b_2$, we obtain
\begin{equation}
\begin{split}
T_\eta(-\bm\rho_j)\ket{\Psi_{\bm\kappa}}
&=(-1)^{n_{j1}n_{j2}}
e^{-2\pi\ii(n_{j1},n_{j2})B\bm\kappa}
\ket{\Psi_{\bm\kappa}}\\
&=(-1)^{n_{j1}n_{j2}}
e^{-2\pi\ii\bm r_j\bm\kappa}
\ket{\Psi_{\bm\kappa}}.
\end{split}
\label{eq:com-qh-character}
\end{equation}
The minus sign in the exponent arises because translating the electrons by $+\bm a_j$ is compensated by translating the quasihole by $-\bm\rho_j$. This determines the part of the electron COM eigenvalue that depends on the quasihole momentum.

The remaining phase follows from applying the theta-function quasiperiodicities to Eq.~\eqref{eq:com-compensation}. The shift of the COM argument by $U_{j1}L_1+U_{j2}L_2$ contributes a constant phase and a boundary-twist-dependent phase. Writing the electron boundary twists as $\bm t=(\theta_1,\theta_2)^T/(2\pi)$, the latter is $e^{2\pi\ii(U\bm t)_j}$. Equating the resulting translation eigenvalue with $e^{2\pi\ii\widetilde\kappa_j}$ gives
\begin{equation}
\boxed{
\widetilde{\bm\kappa}
=R\bm\kappa+\bm c^{(0)}+U\bm t\pmod1,
\qquad R=-r,
}
\label{eq:tilted-affine}
\end{equation}
where $\bm c^{(0)}$ contains the twist-independent theta-function and magnetic-translation phases. For completeness, the constant can be written explicitly once the wavefunction gauge and magnetic-translation conventions have been fixed. Using the physical Weyl-symmetrized quasihole translations in Eq.~\eqref{eq:B-eigen}, and denoting $\varepsilon=\Nphi\bmod2$, the offset in this convention is
\begin{equation}
\begin{split}
c_j^{(0)}=\frac12\big[&
mU_{j1}U_{j2}
+\Ne(Q_{j1}+Q_{j2}-Q_{j1}Q_{j2})\\
&+(U_{j1}+U_{j2})(\varepsilon+1)
+n_{j1}n_{j2}
\big]\pmod1.
\end{split}
\label{eq:c-offset}
\end{equation}
The first terms arise from the theta-function quasiperiodicities, while the last is the phase associated with combining the magnetic-cell translations in Eq.~\eqref{eq:com-qh-character}. The numerical value of $\bm c^{(0)}$ depends on the chosen translation operators and momentum origin, including the magnetic-cell convention. When comparing with ED, one must therefore use precisely the same conventions. Equivalently, the offset can be fixed by evaluating the electronic COM translation eigenvalue of one explicitly identified state and checking the result for the remaining states. Matching only the unordered sets of allowed momenta does not fix this correspondence.

Finally, Eq.~\eqref{eq:tilted-affine} must remain valid when the electron boundary twists are varied. Since the quasihole carries charge magnitude $1/m$ of the electron charge, the changes in the momentum labels are
\begin{equation}
\delta\bm\kappa=\frac1mM^{-1}\delta\bm t,
\qquad
\delta\widetilde{\bm\kappa}
=\Ne M^{-1}\delta\bm t.
\label{eq:flux-shifts}
\end{equation}
Here $\bm t$ already contains the factor $1/(2\pi)$. These two relations are consistent with Eq.~\eqref{eq:tilted-affine}, because
\begin{equation}
\begin{split}
\frac1m RM^{-1}+U
&=\frac{-M^{-1}-Q+J+Q}{m}\\
&=\frac{\Nphi-1}{m}M^{-1}
=\Ne M^{-1}.
\end{split}
\label{eq:twist-identity}
\end{equation}

\subsection{Example: tilted cluster with \texorpdfstring{$\Nphi=16$}{Nphi=16} flux quanta}
\label{subsec:worked-cluster}

We illustrate the construction for the tilted square cluster with $\Nphi=16$, $\Ne=5$, and $m=3$. In the original one-flux primitive basis, the torus matrix is
\begin{equation}
M=\begin{pmatrix}4&2\\-4&2\end{pmatrix}.
\end{equation}
The torus periods are therefore $\bm L_1=4\bm a_1+2\bm a_2$ and $\bm L_2=-4\bm a_1+2\bm a_2$.

Following the construction above, we first calculate
\begin{equation}
J=\Nphi M^{-1}
=\begin{pmatrix}2&-2\\4&4\end{pmatrix},
\qquad
Q=-\operatorname{mod}(J,3)
=\begin{pmatrix}-2&-1\\-1&-1\end{pmatrix}.
\end{equation}
The corresponding compensating quasihole displacements and integer winding numbers are
\begin{equation}
r=I+QM
=\begin{pmatrix}-3&-6\\0&-3\end{pmatrix},
\qquad
U=\frac{J+Q}{3}
=\begin{pmatrix}0&-1\\1&1\end{pmatrix}.
\end{equation}
Thus the momentum map is determined by
\begin{equation}
M=\begin{pmatrix}4&2\\-4&2\end{pmatrix},
\qquad
R=-r=\begin{pmatrix}3&6\\0&3\end{pmatrix},
\qquad
U=\begin{pmatrix}0&-1\\1&1\end{pmatrix}.
\label{eq:M16-example}
\end{equation}

One compatible unfolded quasihole magnetic cell is
\begin{equation}
B=\begin{pmatrix}1&2\\0&3\end{pmatrix},
\qquad \det B=3.
\end{equation}
For this choice, the compensating displacements have magnetic-cell coordinates
\begin{equation}
rB^{-1}
=\begin{pmatrix}-3&0\\0&-1\end{pmatrix}.
\end{equation}
Using the translation convention specified above, Eq.~\eqref{eq:c-offset} gives $\bm c^{(0)}=(0,0)$. The relation between electron COM and quasihole momentum therefore becomes
\begin{equation}
\boxed{
\begin{aligned}
\widetilde\kappa_1
&=3\kappa_1+6\kappa_2-t_2,\\
\widetilde\kappa_2
&=3\kappa_2+t_1+t_2,
\end{aligned}
\qquad \pmod1.
}
\label{eq:M16-momentum-map}
\end{equation}
At zero twists, for example, the allowed quasihole momentum $(\kappa_1,\kappa_2)=(1/8,1/4)$ corresponds to the electronic COM momentum $(\widetilde\kappa_1,\widetilde\kappa_2)=(7/8,3/4)$. Equation~\eqref{eq:M16-momentum-map} determines this correspondence state by state, rather than merely matching the two momentum grids.

We now enlarge the electronic unit cell by taking $\bm A_1=2\bm a_1$ and $\bm A_2=\bm a_2$, and corresponding reciprocal lattice vectors $\bm G_1 = \bm g_1/2$ and $\bm G_2 = \bm g_2$, where $\bm g_1$ and $\bm g_2$ are the reciprocal lattice vectors of the original one-flux lattice. The same physical torus then contains eight two-flux unit cells, and its matrix in the enlarged basis is
\begin{equation}
M_{\rm f}
=M\begin{pmatrix}1/2&0\\0&1\end{pmatrix}
=\begin{pmatrix}2&2\\-2&2\end{pmatrix},
\qquad \det M_{\rm f}=8.
\end{equation}
The corresponding folded electronic COM momentum is $\widetilde{\bm\kappa}_{\rm f}
=(2\widetilde\kappa_1,\widetilde\kappa_2)^T$ modulo integers. Thus each folded COM sector contains two states that were distinguished by translations of the original one-flux lattice.

For the quasihole construction, we use the covering torus with periods $(\bm L_1,m\bm L_2)$. Its matrix in the original primitive basis is
\begin{equation}
P=\begin{pmatrix}4&2\\-12&6\end{pmatrix},
\end{equation}
whereas in the enlarged moir\'e basis it is
\begin{equation}
P_{\rm f}
=P\begin{pmatrix}1/2&0\\0&1\end{pmatrix}
=\begin{pmatrix}2&2\\-6&6\end{pmatrix}.
\end{equation}

We choose the folded quasihole magnetic-cell matrix $B_{\rm f}$ in the enlarged moir\'e basis such that $P_{\rm f}=H_{\rm f}B_{\rm f}$, with both matrices integer-valued. One convenient choice is
\begin{equation}
B_{\rm f}
=\begin{pmatrix}1&1\\-1&2\end{pmatrix}\Rightarrow \bm A_1^{\rm qh}=\bm A_1+\bm A_2, \quad \bm A_2^{\rm qh}=-\bm A_1+2\bm A_2,
\qquad \det B_{\rm f}=3.
\end{equation}
The same physical magnetic-cell vectors, expressed in the original one-flux primitive basis, have the coordinate matrix
\begin{equation}
B_{\rm f}\begin{pmatrix}2&0\\0&1\end{pmatrix}
=\begin{pmatrix}2&1\\-2&2\end{pmatrix},
\end{equation}
whose determinant is six. Thus the magnetic cell contains three enlarged moir\'e cells, or six original one-flux cells, corresponding to two effective quasihole flux quanta.

The covering-torus factorization is
\begin{equation}
P_{\rm f}
=\underbrace{\begin{pmatrix}
2&0\\-2&4
\end{pmatrix}}_{H_{\rm f}}B_{\rm f},
\qquad
P=H_{\rm f}B_{\rm f}
\begin{pmatrix}2&0\\0&1\end{pmatrix},
\qquad
\det H_{\rm f}=8.
\label{eq:tilted-fold-example}
\end{equation}
The covering torus therefore contains eight folded magnetic cells. Each folded momentum sector contains two quasihole states, giving the expected total of $16$ quasihole zero modes.

The quasihole momenta are defined by the eigenvalues of translations through the rows of $B_{\rm f}$,
\begin{equation}
\bm u_{\rm f}=B_{\rm f}\bm\kappa_{\rm f},
\qquad
\bk=2\pi(B_{\rm f}A)^{-1}\bm u_{\rm f},
\label{eq:folded-momentum-example}
\end{equation}
where $A$ is the matrix with rows $\bm A_1,\bm A_2$, and $\bm\kappa_{\rm f}$ denotes momentum components in the corresponding moir\'e reciprocal basis. For the present choice of $B_{\rm f}$, the reciprocal quasihole magnetic-cell vectors are
\begin{equation}
\begin{split}
\bm{G}_1^{\rm qh}
&=\frac23\bm G_1+\frac13\bm G_2,\\
\bm{G}_2^{\rm qh}
&=-\frac13\bm G_1+\frac13\bm G_2,
\end{split}
\label{eq:tilted-magnetic-recip}
\end{equation}
so that $\bk=u_{{\rm f},1}\bm{G}_1^{\rm qh}+
u_{{\rm f},2}\bm{G}_2^{\rm qh} = \kappa_{\rm f,1}\bm G_1+\kappa_{\rm f,2}\bm G_2$. The corresponding magnetic BZ is an oblique parallelogram.

The integer factorization is important. A choice of magnetic-cell vectors $(2\bm a_1,3\bm a_2)$ would give $B_{\rm f}=\operatorname{diag}(1,3)$ in the enlarged basis. For this tilted torus, however, $P_{\rm f}B_{\rm f}^{-1}$ is not an integer matrix, so these magnetic-cell vectors are not compatible with the finite covering-torus periods. The construction above retains the actual tilted geometry and its magnetic boundary conditions.

\section{Two-flux moir\'e folding, the quasihole Hamiltonian, and the splitting of anyon bands into two subbands}
\label{sec:folding}
\subsection{Construction of the flat-band wavefunctions
and the modulation \texorpdfstring{$h(\br)$}{h(r)}}
\label{subsec:h-construction}

Here we summarize the exact construction of the two sublattice-polarized flat-band wavefunctions and their relation to a reference LLL with two flux quanta per moir\'e unit cell, following Ref.~\cite{sarkar2025unconventional}.

We consider the quadratic-band-crossing model
\begin{equation}
\mathcal D(\br)
=
-4\bar\partial_z^2+\widetilde A(\br),
\qquad z=x+\ii y,
\end{equation}
where $\widetilde A(\br)$ is the moir\'e-periodic strain field. At a magic strain configuration, the $\Gamma$-point wavefunction satisfies
\begin{equation}
\mathcal D\psi_\Gamma(\br)=0
\end{equation}
and has two zeros at the positions $\br_0^{(1)}$ and $\br_0^{(2)}$ that allow the construction of two independent flat-band Bloch wavefunctions.

For the hexagonal model considered in the main text, these positions are
\begin{equation}
\br_0^{(1)}
=
\frac{\bm A_1+\bm A_2}{3},
\qquad
\br_0^{(2)}
=
\frac{2(\bm A_1+\bm A_2)}{3},
\label{eq:h-generating-zeros}
\end{equation}
where $\bm A_1,\bm A_2$ are the primitive moir\'e lattice vectors.

Since the kinetic operator contains only $\bar\partial_z$, multiplying $\psi_\Gamma(\br)$ by a holomorphic function $f(z)$ gives another zero mode wherever $f$ is nonsingular:
\begin{equation}
\mathcal D[f(z)\psi_\Gamma(\br)]
=
f(z)\mathcal D\psi_\Gamma(\br)=0.
\end{equation}
To obtain Bloch wavefunctions at arbitrary momentum, we construct holomorphic functions that satisfy the required Bloch boundary conditions. Such functions may have poles at $\br_0^{(1)}$ and $\br_0^{(2)}$, which are canceled by the corresponding zeros of $\psi_\Gamma(\br)$.

Let $\bm G_i$ be the reciprocal vectors, $\bm G_i\cdot\bm A_j=2\pi\delta_{ij}$. We introduce the complex numbers
\begin{equation}
a_j=A_{jx}+\ii A_{jy},
\qquad
b_2=G_{2x}+\ii G_{2y},
\qquad
\tau=\frac{a_2}{a_1},
\end{equation}
and define
\begin{equation}
\xi_\mu(z)
=
\frac{z-z_\mu}{a_1},
\qquad
z_\mu=r_{0x}^{(\mu)}+\ii r_{0y}^{(\mu)}.
\end{equation}
We use the odd Jacobi theta function $\vartheta(w,\tau)$, with quasiperiodicities
\begin{equation}
\begin{split}
\vartheta(w+1,\tau)&=-\vartheta(w,\tau),\\
\vartheta(w+\tau,\tau)
&=-e^{-2\pi\ii w-\pi\ii\tau}\vartheta(w,\tau).
\end{split}
\end{equation}

Writing $p=p_x+\ii p_y$ for the complex electronic Bloch momentum, two independent holomorphic functions can be chosen as
\begin{equation}
\boxed{
f_{\bp}^{(\ell)}(z)
=
e^{\ii(\bp\cdot\bm A_1)z/a_1}
\prod_{\mu=1}^{2}
\frac{
\vartheta\!\left(
\xi_\mu(z)
-\frac{p+(\ell-1)b_2}{2b_2},
\tau\right)
}{
\vartheta(\xi_\mu(z),\tau)
},
\qquad \ell=1,2.
}
\label{eq:h-holomorphic-functions}
\end{equation}
Here $\bp$ labels electronic single-particle momentum, distinct from the quasihole momentum used elsewhere in the paper.

These functions satisfy the appropriate Bloch boundary conditions. Their poles at the two chosen positions are canceled by the zeros of $\psi_\Gamma$, giving the exact zero-energy wavefunctions
\begin{equation}
\boxed{
\psi_{\bp}^{(\ell)}(\br)
=
f_{\bp}^{(\ell)}(z)\psi_\Gamma(\br),
\qquad \ell=1,2.
}
\label{eq:h-exact-flatbands}
\end{equation}
The two wavefunctions span the sublattice-polarized flat-band space at each momentum. They need not be orthogonal, and a smooth orthonormal basis can be obtained from their overlap matrix.

We now show that both wavefunctions can be expressed using a common reference LLL. Choose a uniform reference magnetic field with two flux quanta per moir\'e unit cell, so that
\begin{equation}
A_{\rm M}
=
\hat{\bm z}\cdot(\bm A_1\times\bm A_2)
=
4\pi\ell_B^2.
\end{equation}
Let $\hat{\bm G}_2=\bm G_2/|\bm G_2|$. In a magnetic gauge chosen to place the two zeros of the reference LLL at $\br_0^{(1)}$ and $\br_0^{(2)}$, its folded Bloch wavefunctions can be written as
\begin{equation}
\boxed{
\begin{aligned}
\psi_{\bp,\ell}^{\rm LLL}(\br)
={}&
e^{\ii(\bp\cdot\bm A_1)z/a_1}
\prod_{\mu=1}^{2}
\vartheta\!\left(
\xi_\mu(z)
-\frac{p+(\ell-1)b_2}{2b_2},
\tau
\right)\\
&\times
\exp\!\left[
-\frac{\pi}{A_{\rm M}}
\sum_{\mu=1}^{2}
\left(
\hat{\bm G}_2\cdot
(\br-\br_0^{(\mu)})
\right)^2
\right],
\qquad \ell=1,2.
\end{aligned}
}
\label{eq:h-reference-LLL}
\end{equation}
An overall normalization of these reference wavefunctions is left implicit.

Comparing Eqs.~\eqref{eq:h-exact-flatbands} and \eqref{eq:h-reference-LLL}, we find that the momentum-dependent factors are identical. Their only difference is a common, momentum-independent function $h(\br)$:
\begin{equation}
\boxed{
\psi_{\bp}^{(\ell)}(\br)
=
h(\br)\psi_{\bp,\ell}^{\rm LLL}(\br),
\qquad \ell=1,2.
}
\label{eq:h-LLL-correspondence}
\end{equation}
Explicitly, defining $\psi_{\Gamma,1}^{\rm LLL}(\br)$ by Eq.~\eqref{eq:h-reference-LLL} at $\bp=0$ and $\ell=1$, we obtain
\begin{equation}
\boxed{
h(\br)
=
\frac{\psi_\Gamma(\br)}
{\psi_{\Gamma,1}^{\rm LLL}(\br)}
=
\frac{\psi_\Gamma(\br)}
{\prod_{\mu=1}^{2}
\vartheta(\xi_\mu(z),\tau)}
\exp\!\left[
\frac{\pi}{A_{\rm M}}
\sum_{\mu=1}^{2}
\left(
\hat{\bm G}_2\cdot
(\br-\br_0^{(\mu)})
\right)^2
\right].
}
\label{eq:h-explicit}
\end{equation}
The ratio at the common zeros is understood through its regular limiting value.

Equation~\eqref{eq:h-LLL-correspondence} establishes that the two exactly flat bands on one sublattice are obtained from a single reference LLL with two flux quanta per moir\'e unit cell, multiplied by the same function $h(\br)$. The factor $h(\br)$ carries the magnetic quasiperiodicity opposite to that of the reference LLL, so that the physical electronic Bloch wavefunctions satisfy ordinary moir\'e translations. Its modulus squared is periodic under the primitive moir\'e lattice vectors.

For the interaction calculations, we expand
\begin{equation}
|h(\br)|^2
=
\sum_{\bG}w_{\bG}
e^{\ii\bG\cdot\br},
\label{eq:h-Fourier-expansion}
\end{equation}
where the overall normalization is chosen such that $w_{\bm0}=1$. Accordingly,
\begin{equation}
\boxed{
w_{\bG}
=
\frac{
\displaystyle\int_{\rm M}d^2r\,
|h(\br)|^2e^{-\ii\bG\cdot\br}
}{
\displaystyle\int_{\rm M}d^2r\,
|h(\br)|^2
}.
}
\label{eq:h-Fourier-coefficients}
\end{equation}
All coefficients used in our projected interaction are obtained from the strain-derived $\psi_\Gamma(\br)$ through Eq.~\eqref{eq:h-explicit}.

Finally, the correspondence with a folded LLL does not imply that the physical moir\'e unit cell can be reduced. Although the reference LLL can be represented in a smaller magnetic cell containing one flux quantum, the strain field and the resulting physical wavefunctions retain the primitive moir\'e periodicity. In particular, the modulation $|h(\br)|^2$ does not possess the smaller translation symmetry. Therefore, the two exactly flat electronic bands cannot be unfolded into a single electronic Bloch band of a smaller physical lattice.

The same construction applies to the square-lattice model using its corresponding primitive vectors and the two symmetry-related edge-center zeros of $\psi_\Gamma$, with the appropriate magnetic gauge for the reference LLL.
\subsection{Quasihole Hamiltonian, and the splitting of anyon bands into two subbands}
We now consider a modulation $h(\bm r)$ whose primitive cell contains two flux quanta as was described above. For definiteness, we choose
\begin{equation}
\bm A_1=2\bm a_1,\qquad
\bm A_2=\bm a_2,\qquad
\bm A_1\times\bm A_2=4\pi\ell_B^2,
\label{eq:twoflux-cell}
\end{equation}
where $\bm a_1,\bm a_2$ are the primitive vectors of the original one-flux lattice. Denoting their reciprocal vectors by $\bm g_1,\bm g_2$, with $\bm a_i\cdot\bm g_j=2\pi\delta_{ij}$, the reciprocal vectors of the enlarged moir\'e lattice are $\bm G_1=\bm g_1/2$ and $\bm G_2=\bm g_2$. The enlarged cell halves the electronic BZ and folds the original LLL into two degenerate electronic bands. The full two-band single-particle space still has total Chern number $-1$.

A finite reference torus with matrix $M$ is compatible with the enlarged lattice only if
\begin{equation}
M_{\rm f}
=M\begin{pmatrix}1/2&0\\0&1\end{pmatrix}
\label{eq:moire-M}
\end{equation}
has integer entries. Equivalently, both entries of the first column of $M$ must be even. The number of moir\'e cells is $N_{\rm uc}=\det M_{\rm f}=\Nphi/2$. Electronic COM sectors whose original momenta differ by $\bm G_1$ are consequently identified in the enlarged unit cell.

The quasihole magnetic unit cell contains three moir\'e cells for $m=3$, or more generally $m$ moir\'e cells. Since each moir\'e cell encloses two flux quanta, the corresponding magnetic cell encloses two effective quasihole flux quanta. Its reciprocal area is
\begin{equation}
|\mathrm{aBZ}_{\rm f}|
=\frac{\pi}{m\ell_B^2}.
\label{eq:folded-area}
\end{equation}

To distinguish the two reciprocal-coordinate conventions, we write the same physical quasihole momentum as
\begin{equation}
\bk
=\kappa_1\bm g_1+\kappa_2\bm g_2
=\kappa_{{\rm f},1}\bm G_1
 +\kappa_{{\rm f},2}\bm G_2,
\qquad
\bm\kappa_{\rm f}
=\begin{pmatrix}2\kappa_1\\\kappa_2\end{pmatrix}.
\label{eq:folded-BZ}
\end{equation}
Here $\bm\kappa$ retains its definition from Sec.~\ref{subsec:affine-map}, while $\bm\kappa_{\rm f}$ gives the components of the same wavevector in the moir\'e reciprocal basis.

The shape of the magnetic BZ depends on the choice of commuting quasihole translations. For a magnetic cell aligned with the moir\'e primitive vectors, $B_{\rm f}=\operatorname{diag}(1,m)$, one may use $0\leq\kappa_{{\rm f},1}<1$ and $0\leq\kappa_{{\rm f},2}<1/m$. For a general tilted finite torus, we instead choose a compatible integer magnetic-cell matrix $B_{\rm f}$ satisfying
\begin{equation}
P_{\rm f}=H_{\rm f}B_{\rm f},
\qquad
P_{\rm f}
=\begin{pmatrix}1&0\\0&m\end{pmatrix}M_{\rm f},
\qquad
\det B_{\rm f}=m.
\label{eq:folded-cover-factor}
\end{equation}
The enlarged quasihole translation eigenvalues are then specified by $\bm u_{\rm f}=B_{\rm f}\bm\kappa_{\rm f}$, and the corresponding magnetic BZ is the reciprocal cell of $B_{\rm f}A$. The worked example in Sec.~\ref{subsec:worked-cluster} illustrates this construction explicitly.

The periodic modulation breaks translations by the auxiliary one-flux primitive vector $\bm a_1$, while preserving translations by the enlarged moir\'e vectors $\bm A_j$. The Hamiltonian can therefore couple unfolded electronic COM sectors whose momenta differ by $\bm G_1$. These states belong to the same folded COM sector, and the corresponding two quasihole states can mix. The folding alone reorganizes the zero modes; the interaction projected into each two-dimensional quasihole subspace determines their splitting.

The ideal-band single-particle wavefunctions are obtained by multiplying the two folded LLL orbitals by $h(\bm r)$ and orthonormalizing them. At each folded momentum $\bk$, define their overlap matrix
\begin{equation}
S_{ab}(\bk)
=
\mel{\psi^\text{LLL}_{\bk+\bm Q_a}}
{|h(\bm r)|^2}
{\psi^\text{LLL}_{\bk+\bm Q_b}},
\qquad
\bm Q_0=0,\quad \bm Q_1=\bm G_1.
\label{eq:folded-Gram}
\end{equation}
The two orthonormalized orbitals satisfy
\begin{equation}
\begin{split}
|\psi^h_{\bk,a}\rangle
&=\sum_{b=0}^1
h|\psi^\text{LLL}_{\bk+\bm Q_b}\rangle
[S^{-1/2}(\bk)]_{ba},\\
h|\psi^\text{LLL}_{\bk+\bm Q_a}\rangle
&=\sum_{b=0}^1
|\psi^h_{\bk,b}\rangle
[S^{1/2}(\bk)]_{ba}.
\end{split}
\label{eq:folded-orbitals}
\end{equation}
The second equality determines how multiplication by $\prod_{i=1}^{\Ne}h(\bm r_i)$ acts on a many-electron wavefunction. For example, for two electrons occupying both orbitals at the same folded momentum,
\begin{equation}
\begin{split}
&h(\bm r_1)h(\bm r_2)
\frac{1}{\sqrt{2}}
\begin{vmatrix}
\psi^\text{LLL}_{\bk+\bm Q_0}(\bm r_1)&
\psi^\text{LLL}_{\bk+\bm Q_1}(\bm r_1)\\
\psi^\text{LLL}_{\bk+\bm Q_0}(\bm r_2)&
\psi^\text{LLL}_{\bk+\bm Q_1}(\bm r_2)
\end{vmatrix}\\
&\qquad=
\det[S^{1/2}(\bk)]
\frac{1}{\sqrt{2}}
\begin{vmatrix}
\psi^h_{\bk,0}(\bm r_1)&
\psi^h_{\bk,1}(\bm r_1)\\
\psi^h_{\bk,0}(\bm r_2)&
\psi^h_{\bk,1}(\bm r_2)
\end{vmatrix}.
\end{split}
\end{equation}

More generally, let $|I\rangle_0$ denote a Slater determinant of the reference LLL orbitals and $|J\rangle_h$ a Slater determinant of the orthonormalized ideal-band orbitals. Writing the complete single-particle transformation as the block-diagonal matrix
\begin{equation}
\mathcal S=\bigoplus_{\bk}
S^{1/2}(\bk),
\end{equation}
we obtain
\begin{equation}
\prod_{i=1}^{\Ne}h(\bm r_i)|I\rangle_0
=
\sum_J
\det\!\left[
\mathcal S_{J_p I_q}
\right]_{p,q=1}^{\Ne}
|J\rangle_h,
\label{eq:manybody-h}
\end{equation}
where $I_q$ and $J_p$ denote the occupied orbitals of the two determinants, taken in a fixed order. Thus, if $|\Psi^0\rangle=\sum_I C_I|I\rangle_0$, the coefficients of the modulated many-electron wavefunction are
\begin{equation}
(R_h C)_J
=
\sum_I
\det\!\left[
\mathcal S_{J_p I_q}
\right]_{p,q=1}^{\Ne}
C_I \text{ and }|\Psi^h\rangle=R_h|\Psi^0\rangle=\sum_J (R_h C)_J|J\rangle_h
\label{eq:manybody-map}
\end{equation}
This defines the many-electron transformation $R_h$.

We now use $R_h$ to construct the quasihole zero modes of the modulated ideal-band system. Let $|\Psi^0_{\bk,a}\rangle$, $a=1,2$, be the two orthonormal $V_1$ zero modes in a given folded COM momentum sector of the uniform LLL. Multiplication by $\prod_{i=1}^{\Ne}h(\bm r_i)$ gives
\begin{equation}
|\widetilde\Psi^h_{\bk,a}\rangle
=R_h|\Psi^0_{\bk,a}\rangle,
\label{eq:modulated-zero-modes}
\end{equation}
which are zero modes of the corresponding ideal-band $V_1$ parent Hamiltonian. Since multiplication by $h$ does not preserve the normalization of the wavefunctions, we calculate their overlap matrix
\begin{equation}
\begin{split}
G_{ab}(\bk)
&=\langle\widetilde\Psi^h_{\bk,a}|
\widetilde\Psi^h_{\bk,b}\rangle\\
&=
\left\langle\Psi^0_{\bk,a}\left|
\prod_{i=1}^{\Ne}|h(\bm r_i)|^2
\right|\Psi^0_{\bk,b}\right\rangle.
\end{split}
\label{eq:parent-Gram}
\end{equation}
The orthonormal quasihole states are then
\begin{equation}
|\Psi^h_{\bk,a}\rangle
=
\sum_{b=1}^{2}
|\widetilde\Psi^h_{\bk,b}\rangle
[G^{-1/2}(\bk)]_{ba},
\label{eq:parent-polar}
\end{equation}
which satisfy $\langle\Psi^h_{\bk,a}|
\Psi^h_{\bk,b}\rangle=\delta_{ab}$.
The two states remain in the same folded COM sector because $h(\bm r)$ is periodic under the moir\'e lattice translations.

The quasihole wavefunction in Eq.~\eqref{eq:qh-torus} also contains the factor $g(\eta)$, so that
\begin{equation}
\Psi_\sigma^h(\{\bm r_i\};\eta)
=g(\eta)\prod_{i=1}^{\Ne}h(\bm r_i)
\Psi_\sigma^0(\{\bm r_i\};\eta).
\end{equation}
Since $g(\eta)$ is a nonzero scalar independent of the electron coordinates, it does not change the electronic zero-mode subspace. However, it contributes the factor $|g(\eta)|^2$ to the quasihole-coordinate overlaps and must therefore be retained in the Berry-curvature calculation of Sec.~\ref{sec:chern}.

To obtain the quasihole dispersion, we project the screened Coulomb interaction $H_{\rm sc}$ onto the two orthonormal zero modes $|\Psi^h_{\bk,a}\rangle$ constructed above. The resulting Hamiltonian is
\begin{equation}
H^{(2)}_{ab}(\bk)
=
\mel{\Psi^h_{\bk,a}}
{H_{\rm sc}}
{\Psi^h_{\bk,b}},
\qquad a,b=1,2.
\label{eq:two-state-H}
\end{equation}
Diagonalizing this $2\times2$ matrix,
\begin{equation}
\sum_{b=1}^2
H^{(2)}_{ab}(\bk)v_{\pm,b}
=E_\pm(\bk)v_{\pm,a},
\end{equation}
gives the two quasihole energies $E_\pm$ and the corresponding many-electron wavefunctions
\begin{equation}
|\Phi_{\bk,\pm}\rangle
=
\sum_{a=1}^2
v_{\pm,a}(\bk)
|\Psi^h_{\bk,a}\rangle.
\label{eq:anyon-eigenstates}
\end{equation}
For uniform $h$, magnetic translation symmetry makes the two quasihole energies degenerate. When $h$ has the periodicity of the two-flux moir\'e lattice, the off-diagonal matrix elements of $H^{(2)}$ are generally nonzero, and the two quasihole bands can split. The energies $E_\pm$ are variational energies obtained by restricting $H_{\rm sc}$ to the exact $V_1$ zero-mode subspace; the states $|\Phi_{\bk,\pm}\rangle$ are not in general exact eigenstates of the full screened Coulomb Hamiltonian; we will call $|\Phi_{\bk,\pm}\rangle$ variational quasi-hole states.

\subsection{Variational quasihole states and the screened-Coulomb spectrum}
\label{subsec:coulomb-comparison}

\begin{figure}[t]
\centering
\includegraphics[width=\linewidth]{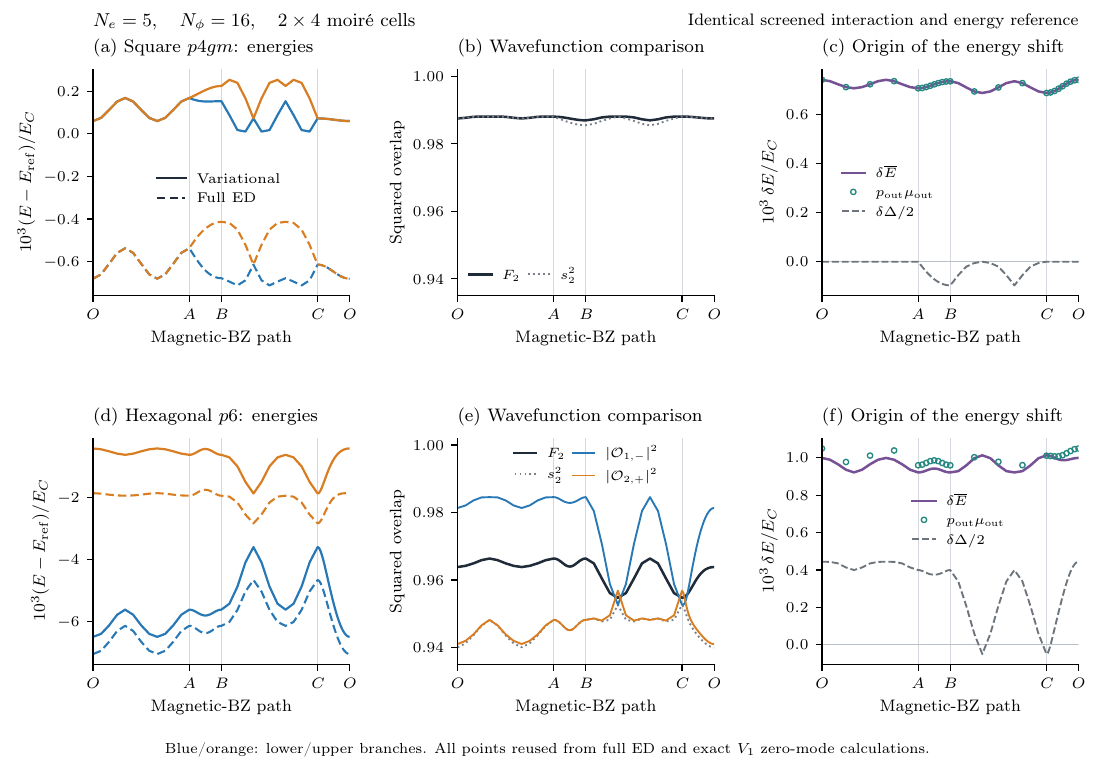}
\caption{Zero-mode projection compared with full screened-Coulomb ED for square $p4gm$ (top) and hexagonal $p6$ (bottom), on aligned $2\times4$ moir\'e clusters with $\Ne=5$ and $\Nphi=16$. (a,d) Variational (solid) and exact (dashed) energies; blue/orange denote lower/upper branches. Both use the same uniform five-electron reference $E_{\rm ref}$, with $E_C=e^2/(\epsilon\ell_B)$. (b,e) Average doublet fidelity $F_2$, minimum projection weight $s_2^2$ [Eq.~\eqref{eq:coulomb-subspace-fidelity}], and individual squared overlaps for the isolated $p6$ branches. (c,f) Centroid shift $\delta\overline E$, half the splitting error $\delta\Delta/2$ [Eq.~\eqref{eq:coulomb-energy-errors}], and the energy-weighted tail $p_{\rm out}\mu_{\rm out}$ [Eq.~\eqref{eq:energy-tail-decomposition}]. The magnetic-BZ path is $O\to A\to B\to C\to O$, with $(\kappa_{{\rm f},1},\kappa_{{\rm f},2})=(0,0),(1/2,0),(1/2,1/6),(0,1/6),(0,0)$. Lines connect the evaluated momenta and serve as guides to the eye.}
\label{fig:variational-coulomb-comparison}
\end{figure}

The zero-mode projection in Eq.~\eqref{eq:two-state-H} is useful if it captures the low-energy Coulomb subspace and the dispersion of its two quasihole branches. These two aspects of the approximation need not have the same accuracy. We examine them by comparing Eq.~\eqref{eq:anyon-eigenstates} with full ED in the electronic Hilbert space formed from both bands. We use the double gated screened interaction
\begin{equation}
V_{\rm sc}(q)=\frac{2\pi e^2}{\epsilon q}
\tanh(qd_{\rm sc}),\qquad d_{\rm sc}|\bG_1|=2,
\label{eq:screened-coulomb-interaction}
\end{equation}
where $q=|\bm q|$ and $\epsilon$ is the dielectric constant. Thus $d_{\rm sc}/\ell_B=1.12838$ for the square lattice and $1.05008$ for the hexagonal lattice. The comparisons below are for aligned $2\times4$ moir\'e clusters with $\Ne=5$ and $\Nphi=16$. Searching the full Coulomb spectrum by its overlap with the zero-mode space identifies the two lowest states as the quasihole doublet at every sampled momentum. This doublet remains separated from higher Coulomb states.

Let $|\Psi^{\rm ED}_{\bk,n}\rangle$ denote the Coulomb eigenstates, ordered by their energies $E_n^{\rm ED}(\bk)$, and define the electronic overlaps
\begin{equation}
\mathcal O_{n\alpha}(\bk)
=\langle\Psi^{\rm ED}_{\bk,n}|\Phi_{\bk,\alpha}\rangle,
\qquad \alpha=\pm.
\label{eq:coulomb-overlaps}
\end{equation}
The $n=1,2$ block, denoted $\mathcal O^{(2)}$, compares two orthonormal bases of the variational and Coulomb doublets. Its singular values $1\geq s_1\geq s_2\geq0$ are the square roots of the eigenvalues of $\mathcal O^{(2)\dagger}\mathcal O^{(2)}$. Writing $P_0$ and $P_{\rm ED}$ for the projectors onto these two spaces, respectively, gives
\begin{equation}
\begin{split}
F_2(\bk)
&=\frac12\operatorname{Tr}(P_0P_{\rm ED})
=\frac{s_1^2+s_2^2}{2},\\
s_2^2(\bk)
&=\min_{\sum_\alpha|c_\alpha|^2=1}
\left\|P_{\rm ED}\sum_{\alpha=\pm}
c_\alpha|\Phi_{\bk,\alpha}\rangle\right\|^2.
\end{split}
\label{eq:coulomb-subspace-fidelity}
\end{equation}
Thus $F_2$ measures the average projection into the exact doublet, whereas $s_2^2$ tests its least well represented direction. The latter prevents a good overlap of one state from concealing a poor description of the other. Both quantities are independent of basis rotations within either doublet, which is essential at the protected degeneracies in the $p4gm$ model. In the $p6$ model, where the two anyon bands are fully gapped, overlaps of the ED and variational states corresponding to each band also test the identification of the two energy eigenstates.

For square $p4gm$ and hexagonal $p6$, the minimum sampled $F_2$ is $0.9869$ and $0.9547$, respectively, while the minimum $s_2$ is $0.9927$ and $0.9696$. Even the least well represented
superposition therefore retains approximately $98.55\%$ and
$94.00\%$ of its norm in the exact Coulomb doublet. Together
with spectral isolation, these results support the zero-mode space as a description of the low-energy quasihole degrees of freedom on the tested clusters. In $p6$, the individual squared overlaps remain above $0.9410$; in $p4gm$, the doublet quantities allow the comparison to continue through the band touchings without choosing arbitrary eigenvectors there.

The energies provide a separate test. The variational principle gives $E_-(\bk)\geq E_1^{\rm ED}(\bk)$ and $E_+(\bk)\geq E_2^{\rm ED}(\bk)$. Define $\delta E_-=E_--E_1^{\rm ED}$ and $\delta E_+=E_+-E_2^{\rm ED}$, and separate their common and differential parts,
\begin{equation}
\delta\overline E=\frac{\delta E_-+\delta E_+}{2},\qquad
\delta\Delta=\delta E_+-\delta E_-.
\label{eq:coulomb-energy-errors}
\end{equation}
The first measures the displacement of the doublet as a whole; the second measures the error in its splitting. The uniform system has a constant common shift. Modulation introduces a momentum dependence and a differential correction: the latter is small for $p4gm$ but more pronounced for $p6$ [Fig.~\ref{fig:variational-coulomb-comparison}(c,f)]. Consequently, a rigid energy shift does not generally recover the exact Coulomb dispersion.

The coexistence of high fidelity and a visible energy shift follows from the energy carried by the small spectral weight above the doublet. Completeness gives
\begin{equation}
E_\alpha(\bk)=\sum_n|\mathcal O_{n\alpha}(\bk)|^2
E_n^{\rm ED}(\bk).
\label{eq:variational-spectral-energy}
\end{equation}
Set $W_n=\sum_\alpha|\mathcal O_{n\alpha}|^2$ and $\overline E^{\rm ED}=(E_1^{\rm ED}+E_2^{\rm ED})/2$. The total weight above the exact doublet, averaged over the two trial states, is $p_{\rm out}=1-F_2$. Its contribution to the centroid shift is
\begin{equation}
\begin{split}
\delta\overline E
&=\frac12\sum_{n=1}^{2}W_n
(E_n^{\rm ED}-\overline E^{\rm ED})
+p_{\rm out}\mu_{\rm out},\\
p_{\rm out}\mu_{\rm out}
&=\frac12\sum_{n\geq3}W_n
(E_n^{\rm ED}-\overline E^{\rm ED}).
\end{split}
\label{eq:energy-tail-decomposition}
\end{equation}
Here $\mu_{\rm out}$ is the mean excitation energy of this spectral tail relative to the doublet centroid. The magnetic-BZ
average of $p_{\rm out}$ is only $1.22\%$ for $p4gm$ and
$3.73\%$ for $p6$, but the higher excitation energies give this
small weight an appreciable energetic cost. The agreement of $p_{\rm out}\mu_{\rm out}$ with the centroid shift in Fig.~\ref{fig:variational-coulomb-comparison}(c,f) identifies this tail as the dominant contribution. The remaining term measures unequal weights within the exact doublet and vanishes when $W_1=W_2$.

The action of $H_{\rm sc}$ outside the $V_1$ pseudopotential zero-mode space is quantified independently by
\begin{equation}
\begin{split}
r_\alpha^2(\bk)
&=\big\|(1-P_0)H_{\rm sc}|\Phi_{\bk,\alpha}\rangle\big\|^2\\
&=\sum_n|\mathcal O_{n\alpha}|^2
\big(E_n^{\rm ED}-E_\alpha\big)^2.
\end{split}
\label{eq:variational-residual}
\end{equation}
Since $P_0H_{\rm sc}|\Phi_{\bk,\alpha}\rangle
=E_\alpha|\Phi_{\bk,\alpha}\rangle$, this residual is the energy variance of the trial state. It measures coupling omitted by the projection, rather than an error in solving the $2\times2$ Hamiltonian. At a degeneracy we use the invariant sum $\sum_\alpha r_\alpha^2$. The exact $V_1$ zero-mode condition does not require this Coulomb residual to vanish: the two Hamiltonians have different eigenstates. The $V_1$ parent gap, the splitting within the Coulomb doublet, and its isolation from higher Coulomb states therefore serve distinct purposes.

\section{Composite Glide--time-reversal symmetry and protected quasihole degeneracies}
\label{sec:glide}

We consider ideal Chern bands with magnetic composite glide--time-reversal symmetry. symmetry. Let $O_y(x,y)=(x,-y)$ and $O_x(x,y)=(-x,y)$ denote reflections about the $x$ and $y$ axes, respectively. For a square moir\'e lattice with primitive vectors $\bm A_1=(A,0)$ and $\bm A_2=(0,A)$, the corresponding glide operations involve the fractional translations
\begin{equation}
\bm d_y=\frac{\bm A_1-\bm A_2}{2},
\qquad
\bm d_x=\frac{\bm A_1+\bm A_2}{2}.
\label{eq:glide-def}
\end{equation}
These satisfy $\bm d_y+O_y\bm d_y=\bm A_1$ and $\bm d_x+O_x\bm d_x=\bm A_2$.

We first consider an ideal Chern band without an external magnetic field. As discussed in Sec.~\ref{sec:zero-modes}, the internal spinor $\chi(\bm r)$ cancels the magnetic quasiperiodicity of the scalar LLL wavefunction, so the physical electrons transform under ordinary lattice translations. Let $\mathsf t(\bm d)$ denote an ordinary electronic translation and $\Xi_j$ the antiunitary reflection-times-time-reversal operation. For a spinless realization with $\Xi_j^2=1$, the glide operators are $\mathcal G_j=\mathsf t(\bm d_j)\Xi_j$. Using
\begin{equation}
\Xi_j\mathsf t(\bm d)\Xi_j^{-1}
=\mathsf t(O_j\bm d),
\end{equation}
we find
\begin{align}
\mathcal G_y^2
&=\mathsf t(\bm d_y)
\mathsf t(O_y\bm d_y)
=\mathsf t(\bm A_1),
\nonumber\\
\mathcal G_x^2
&=\mathsf t(\bm d_x)
\mathsf t(O_x\bm d_x)
=\mathsf t(\bm A_2).
\label{eq:glide-physical-square}
\end{align}
Applying the glides to all $\Ne$ electrons gives
\begin{equation}
\boxed{
(\mathcal G_y^{(\Ne)})^2
=\mathsf t_{\rm COM}(\bm A_1),
\qquad
(\mathcal G_x^{(\Ne)})^2
=\mathsf t_{\rm COM}(\bm A_2).
}
\label{eq:glide-physical-manybody}
\end{equation}
For a many-electron state with physical COM momentum $\bm K$,
\begin{equation}
\mathsf t_{\rm COM}(\bm A_j)
\ket{\Psi_{\bm K}}
=e^{\ii\bm K\cdot\bm A_j}
\ket{\Psi_{\bm K}}.
\end{equation}
The antiunitary glide $\mathcal G_y$ maps $(K_x,K_y)$ to $(-K_x,K_y)$, while $\mathcal G_x$ maps it to $(K_x,-K_y)$. On the corresponding BZ boundaries, $K_x=\pi/A$ and $K_y=\pi/A$, respectively, the glide preserves the momentum sector and its square is $-1$. Every energy eigenstate on these lines therefore has an orthogonal degenerate partner.

We next consider an ideal band in an external magnetic field. In this case the electronic translations are magnetic translations, denoted by $T_e(\bm d)$. Reflection reverses the magnetic field, as does complex conjugation, so their combination $\Xi_j$ preserves the magnetic problem. The antiunitary glides are $\mathcal G_j=T_e(\bm d_j)\Xi_j$. Using $\Xi_jT_e(\bm d)\Xi_j^{-1}=T_e(O_j\bm d)$ and the magnetic translation composition law, we obtain
\begin{align}
\mathcal G_y^2
&=e^{\ii\bm d_y\times O_y\bm d_y/(2\ell_B^2)}
T_e(\bm A_1)
=-T_e(\bm A_1),
\nonumber\\
\mathcal G_x^2
&=e^{\ii\bm d_x\times O_x\bm d_x/(2\ell_B^2)}
T_e(\bm A_2)
=-T_e(\bm A_2).
\label{eq:glide-square}
\end{align}
Here we used
\begin{equation}
\bm d_y\times O_y\bm d_y
=\bm d_x\times O_x\bm d_x
=\frac{\bm A_1\times\bm A_2}{2}
=2\pi\ell_B^2.
\end{equation}
Thus the additional minus sign arises from the magnetic translation phase associated with the two-flux moir\'e unit cell. For $\Ne$ electrons, the corresponding relations become
\begin{equation}
\boxed{
\begin{aligned}
(\mathcal G_y^{(\Ne)})^2
&=(-1)^{\Ne}U_{\rm COM}(\bm A_1),\\
(\mathcal G_x^{(\Ne)})^2
&=(-1)^{\Ne}U_{\rm COM}(\bm A_2),
\end{aligned}
}
\label{eq:glide-manybody}
\end{equation}
where $U_{\rm COM}(\bm A_j)$ is the many-electron magnetic translation.

For the one-quasihole systems considered here, $\Nphi=3\Ne+1$ is even, so $\Ne$ is odd. Equation~\eqref{eq:glide-manybody} then implies that a glide squares to $-1$ on a glide-invariant COM sector where the corresponding magnetic translation has eigenvalue $+1$. The protecting condition differs in appearance from the ordinary-translation result, but the absolute momentum assigned to a magnetic translation character depends on the constant phase convention of the translation operator. For example, the redefinition $U_{\rm COM}(\bm A_j)\to-U_{\rm COM}(\bm A_j)$ shifts the corresponding momentum origin by half a reciprocal lattice vector and changes the written glide-square relation by the same sign. The degeneracy condition itself, $(\mathcal G_j^{(\Ne)})^2=-1$, is unchanged.

In either case, an antiunitary glide satisfying $\mathcal G_j^2=-1$ within an invariant momentum sector enforces a twofold degeneracy. Indeed, if $\mathcal G_j\ket{\Psi}=c\ket{\Psi}$, then antiunitarity gives $\mathcal G_j^2\ket{\Psi}=|c|^2\ket{\Psi}$, contradicting $\mathcal G_j^2=-1$. Since the glide commutes with the Hamiltonian, $\ket{\Psi}$ and $\mathcal G_j\ket{\Psi}$ are orthogonal states with the same energy.

\subsection{Positions of the degeneracy lines in the magnetic anyon BZ}
\label{subsec:glide-lines}

We now determine how the glide-protected electronic COM momenta are represented in the quasihole magnetic BZ. We focus on the zero-field realization, for which the physical electronic translations are ordinary translations and the protected COM momenta lie at the BZ boundaries.

Consider a finite torus with two moir\'e cells along $\bm A_1$ and four along $\bm A_2$. We label its electronic COM sectors by $K_1=0,1$ and $K_2=0,1,2,3$, such that
\begin{equation}
\begin{split}
\mathsf t_{\rm COM}(\bm A_1)
\ket{\Psi_{K_1,K_2}}
&=e^{\ii\pi K_1}\ket{\Psi_{K_1,K_2}},\\
\mathsf t_{\rm COM}(\bm A_2)
\ket{\Psi_{K_1,K_2}}
&=e^{\ii\pi K_2/2}\ket{\Psi_{K_1,K_2}}.
\end{split}
\end{equation}
Equation~\eqref{eq:glide-physical-manybody} then gives the protected sectors
\begin{equation}
\boxed{
K_1=1\quad\text{or}\quad K_2=2.
}
\label{eq:glide-protected-com}
\end{equation}
These are the boundaries of the physical electronic COM BZ, including their intersection.

To express the same states in the magnetic anyon BZ, we use the COM-to-anyon momentum correspondence derived in Sec.~\ref{subsec:affine-map}. For the present torus, in the theta-function convention used above, the quasihole momentum $\bk=\kappa_1\bm G_1+\kappa_2\bm G_2$ is given in analytic sector $\sigma=1$ by
\begin{equation}
\boxed{
\begin{aligned}
\kappa_1
&=\frac{K_1}{2}
+\frac{\theta_1}{12\pi}
+\frac12,\\
\kappa_2
&=\frac{K_2}{12}
+\frac{\theta_2}{24\pi}
+\frac16.
\end{aligned}
}
\label{eq:p4gm-momentum}
\end{equation}
Here $\bm G_1,\bm G_2$ are the reciprocal vectors of the physical two-flux moir\'e lattice. The constant terms specify the momentum origin in this convention and must be retained when relating electronic and quasihole momenta.

At zero twists, the sectors protected by $\mathcal G_y$ have $K_1=1$, giving
\begin{equation}
\kappa_1=\frac12+\frac12
=0\pmod1.
\end{equation}
The other analytic topological sectors are related by quasihole magnetic translations, which shift $\kappa_1$ by $2(\sigma-1)/3$ modulo one. The vertical degeneracy lines are therefore
\begin{equation}
\boxed{
\kappa_1=0,\quad\frac13,\quad\frac23
\pmod1.
}
\label{eq:vertical-lines}
\end{equation}
The complementary sectors $K_1=0$ occur at
\begin{equation}
\kappa_1=\frac12,\quad\frac16,\quad\frac56
\pmod1,
\label{eq:unprotected-lines}
\end{equation}
where $\mathcal G_y$ alone does not enforce a degeneracy.

Similarly, the sectors protected by $\mathcal G_x$ have $K_2=2$, so
\begin{equation}
\kappa_2=\frac2{12}+\frac16
=\frac13\equiv0\pmod{1/3}.
\end{equation}
Thus the horizontal degeneracy line is
\begin{equation}
\boxed{
\kappa_2=0\pmod{1/3}.
}
\label{eq:horizontal-line}
\end{equation}
The two sets of lines intersect at momenta where both glide symmetries can act within the same quasihole subspace.

For a tilted torus, the glide operations must preserve the finite boundary conditions and the relevant momentum sectors. The protected momenta are then obtained by combining the glide-square conditions with the general COM-to-anyon correspondence in Sec.~\ref{subsec:affine-map}, rather than using the aligned-torus momentum formulas.

Since the two quasihole bands touch along the glide-protected lines, they do not have separately defined Chern numbers in the symmetric model. Their combined rank-two subspace can nevertheless remain isolated from higher quasihole excitations and carries a well-defined Chern number. Breaking the protecting glide symmetries can separate the two bands, in which case their individual Chern numbers can be calculated.


\section{Berry curvature and Chern numbers from ED quasihole states}
\label{sec:chern}

We now calculate the Berry curvature and Chern numbers of the quasihole bands obtained by projecting the screened Coulomb interaction into the $V_1$ pseudopotential zero-mode space. Recall that the two zero modes $|\Psi^h_{\bk,a}\rangle$, $a=1,2$, provide a basis at each magnetic anyon momentum $\bk$. Diagonalizing the projected screened Coulomb interaction gives the eigenstates defined in Eq.~\eqref{eq:anyon-eigenstates}, \[
|\Phi_{\bk,\pm}\rangle
=
\sum_{a=1}^2
v_{\pm,a}(\bk)|\Psi^h_{\bk,a}\rangle.
\] These are the many-electron states whose quasihole-band topology we wish to determine. In the following, $|\Phi_{\bk,a}\rangle$, with $a=\pm$, denotes a normalized eigenstate of the projected screened interaction, whereas $\varphi_{a,\bk}(\eta)$ denotes its corresponding scalar wavefunction in the quasihole coordinate $\eta$. In the uniform LLL, the two folded quasihole states are degenerate, and only the Chern number of their combined subspace is unambiguously defined. In the modulated system, the screened interaction can separate them into isolated bands or leave symmetry-protected band crossings.

To calculate the quasihole-band Chern numbers, we construct the overlaps of the cell-periodic quasihole wavefunctions at neighboring momenta in the magnetic anyon BZ. We first define these overlaps and distinguish them from electronic many-body overlaps under boundary-twist variations. We then evaluate them analytically using the electronic eigenstates $|\Phi_{\bk,a}\rangle$.

For each quasihole position $\eta$, the localized Laughlin quasihole wavefunction in Eq.~\eqref{eq:qh-torus} defines a state $|\psi_\sigma(\eta)\rangle$ in the electronic many-body Hilbert space. We define the quasihole-coordinate amplitude of $|\Phi_{\bk,a}\rangle$ by
\begin{equation}
\mathcal A_{a,\bk}(\eta)
=
\langle\psi_\sigma(\eta)|\Phi_{\bk,a}\rangle_e.
\label{eq:ed-coordinate}
\end{equation}
Here the subscript $e$ denotes the inner product over the electron coordinates and internal components. Unlike $|\psi_\sigma(\eta)\rangle$, $\mathcal A_{a,\bk}(\eta)$ is a scalar function of the quasihole position. It is linear in the many-electron state $|\Phi_{\bk,a}\rangle$, just as the usual wavefunction $\langle\bm r|\Psi\rangle$ is linear in $|\Psi\rangle$.

We normalize this scalar wavefunction on the quasihole covering torus $\mathcal C$,
\begin{equation}
\boxed{
\varphi_{a,\bk}(\eta)
=
\frac{\mathcal A_{a,\bk}(\eta)}
{\left[
\int_{\mathcal C}d^2\eta\,
|\mathcal A_{a,\bk}(\eta)|^2
\right]^{1/2}}.
}
\end{equation}
We will sometimes refer to it as quasihole-coordinate wavefunction and denote its cell-periodic part by
\begin{equation}
u_{a,\bk}(\eta)
=
e^{-\ii\bk\cdot\be}\varphi_{a,\bk}(\eta).
\label{eq:anyon-periodic}
\end{equation}
The momentum $\bk$ is defined through the magnetic translation eigenvalues of the quasihole-coordinate wavefunction. The same quasihole momentum label is used for the electronic eigenstate $|\Phi_{\bk,a}\rangle$ and its coordinate representation $\varphi_{a,\bk}(\eta)$, following the COM-to-anyon correspondence established in Sec.~\ref{subsec:affine-map}.

For two neighboring momenta $\bk$ and $\bk'=\bk+\bq$, the overlap required for the Berry curvature is
\begin{equation}
\begin{split}
\mathcal L_{ab}(\bk',\bk)
&=
\langle u_{a,\bk'}|u_{b,\bk}\rangle_\eta\\
&=
\int_{\mathcal C}d^2\eta\,
\varphi_{a,\bk'}^*(\eta)
e^{\ii\bq\cdot\be}
\varphi_{b,\bk}(\eta).
\end{split}
\label{eq:anyon-link}
\end{equation}

This overlap should be distinguished from one obtained by directly comparing the electronic many-body states $|\Phi_{\bk,a}\rangle$ under boundary-twist variations. Indeed, the localized quasihole state $|\psi_\sigma(\eta)\rangle$ itself depends on the electronic boundary twists. Consequently, when $\bk$ is varied by changing the twists, both $|\Phi_{\bk,a}\rangle$ and $|\psi_\sigma(\eta)\rangle$ change. In a common electronic gauge, differentiating Eq.~\eqref{eq:ed-coordinate} gives
\begin{equation}
\begin{split}
\partial_{k_\mu}\mathcal A_{a,\bk}(\eta)
={}&
\langle\partial_{k_\mu}\psi_\sigma(\eta)|
\Phi_{\bk,a}\rangle_e\\
&+
\langle\psi_\sigma(\eta)|
\partial_{k_\mu}\Phi_{\bk,a}\rangle_e.
\end{split}
\label{eq:coordinate-derivative}
\end{equation}
The first term accounts for the variation of the localized Laughlin quasihole wavefunction with the electronic twists. The normalization of $\varphi_{a,\bk}$ and the Bloch phase in Eq.~\eqref{eq:anyon-periodic} also contribute to its Berry connection. Thus the Berry curvature defined by Eq.~\eqref{eq:anyon-link} is not generally equal to the curvature obtained from $\ii\langle\Phi_{\bk,a}|
\partial_{\bk}\Phi_{\bk,a}\rangle_e$.
The former describes the Bloch-band geometry of the quasihole, whereas the latter, with the proper electronic gauge and boundary identifications, probes the electronic many-body response to flux insertion. We return to this distinction below.

Before normalization, the integrand in Eq.~\eqref{eq:anyon-link} is proportional to
\begin{equation}
\langle\Phi_{\bk',a}|
\psi_\sigma(\eta)\rangle_e\,
e^{\ii\bq\cdot\be}\,
\langle\psi_\sigma(\eta)|
\Phi_{\bk,b}\rangle_e.
\end{equation}
The electronic inner products have already been taken; the remaining integration is over the quasihole coordinate alone.

For a finite torus, the allowed quasihole momenta form a discrete set at fixed electronic boundary twists $\bm\theta=(\theta_1,\theta_2)^T$. To calculate the overlaps at sufficiently close momenta, we vary these twists. In the present coordinate convention, the quasihole translation characters are complex conjugates of the holomorphic characters used in Sec.~\ref{subsec:affine-map}. Consequently, expressing the momentum in the original one-flux reciprocal basis gives
\begin{equation}
\delta\bm\kappa
=
-\frac{1}{2\pi m}M^{-1}\delta\bm\theta.
\label{eq:chern-flux-shift}
\end{equation}
Varying $\bm\theta$ therefore allows us to construct a finer quasihole momentum mesh. The two states in Eq.~\eqref{eq:anyon-link} generally have different electronic boundary conditions, which must be retained when evaluating their overlap. The twists are used to access neighboring quasihole momenta; the Berry curvature is calculated over the magnetic anyon BZ rather than over the electronic boundary-twist torus.

Evaluation of $V_1$ pseudopotential zero modes using ED and subsequent projection of screened Coulomb interaction to obtain the many-electron eigenstates provide the many-electron eigenstates through their coefficients in an orthonormal Slater-determinant basis,
\begin{equation}
|\Phi_{\bk,a}\rangle
=
\sum_I C_{I,a}(\bk)|I\rangle.
\end{equation}
Although the COM-to-anyon correspondence determines the quasihole momentum of each eigenstate, these coefficients do not directly give Eq.~\eqref{eq:anyon-link}. In particular, the operator $e^{\ii\bq\cdot\be}$ acts on the quasihole coordinate, and its matrix elements are not ordinary Slater-determinant overlaps.

We now construct the required matrix elements from the exact Laughlin quasihole wavefunction. We first establish the correspondence between the electronic zero modes and quasihole-coordinate wavefunctions in the uniform LLL. We then evaluate the phase insertion analytically and extend the construction to the modulated ideal band, retaining the factor $g(\eta)$.

\subsection{Expansion of the localized quasihole wavefunction}
\label{subsec:overlap-matrices}

As established in Sec.~\ref{sec:tilted}, fixing one analytic sector $\sigma$ and allowing $\eta$ to range over the covering torus with periods
\begin{equation}
\bm C_1=\bm L_1,\qquad
\bm C_2=m\bm L_2
\end{equation}
incorporates all $m$ analytic sectors. The quasihole has effective magnetic length $\qell^2=m\ell_B^2$, and the cover contains
\begin{equation}
\frac{\bm C_1\times\bm C_2}{2\pi\qell^2}
=
\Nphi
\end{equation}
effective flux quanta. Thus the $\eta$ dependence of the localized uniform Laughlin quasihole wavefunction belongs to an $\Nphi$-dimensional scalar LLL on the covering torus.

We rotate the coordinates so that $\bm C_1=(X,0)$ and $\bm C_2=(V,Y)$, with $Y>0$. Since $XY=2\pi\qell^2\Nphi$, an orthonormal basis of the scalar LLL is
\begin{equation}
\begin{split}
\phi_j^{\alpha,\beta}(x,y)
={}&
\frac{e^{\ii xy/(2\qell^2)}}
{\sqrt{X\sqrt{\pi\qell^2}}}
\sum_{n\in\mathbb Z}
e^{-\ii n\beta}
\exp\!\left[
\frac{\ii\qell^2 V}{2Y}p_{j+n\Nphi}^2
+\ii p_{j+n\Nphi}x
-\frac{(y+\qell^2p_{j+n\Nphi})^2}
{2\qell^2}
\right],\\
&p_{j+n\Nphi}
=
\frac{2\pi(j+n\Nphi)+\alpha}{X},
\qquad j=0,\ldots,\Nphi-1.
\end{split}
\label{eq:oblique-orbitals}
\end{equation}
Here $\alpha=\alpha_\sigma$ and $\beta$ are the covering-torus boundary phases defined in Eq.~\eqref{eq:ab-boundaries}. The quadratic phase proportional to $V/Y$ ensures the correct boundary conditions on the oblique torus. These functions satisfy
\begin{equation}
\int_{\mathcal C}d^2\eta\,
[\phi_i^{\alpha,\beta}(\eta)]^*
\phi_j^{\alpha,\beta}(\eta)
=
\delta_{ij}.
\end{equation}

We now regard the localized uniform quasihole wavefunction $|\psi_\sigma^0(\eta)\rangle$ as a many-electron state for each fixed $\eta$. Since its dependence on $\eta$ lies in the scalar LLL just constructed, we can expand it exactly as
\begin{equation}
\boxed{
|\psi_\sigma^0(\eta)\rangle
=
\sum_{j=0}^{\Nphi-1}
\phi_j^{\alpha,\beta}(\eta)|v_j\rangle.
}
\label{eq:kernel-expand}
\end{equation}
This equation defines the many-electron states $|v_j\rangle$. Using the orthonormality of the scalar functions, we obtain
\begin{equation}
\boxed{
|v_j\rangle
=
\int_{\mathcal C}d^2\eta\,
[\phi_j^{\alpha,\beta}(\eta)]^*
|\psi_\sigma^0(\eta)\rangle.
}
\label{eq:v-definition}
\end{equation}
The states $|v_j\rangle$ are independent of $\eta$ and are generally correlated many-electron states, not individual Slater determinants. Since $|\psi_\sigma^0(\eta)\rangle$ is a $V_1$ zero mode for every $\eta$, Eq.~\eqref{eq:v-definition} gives
\begin{equation}
H_{V_1}^0|v_j\rangle=0.
\end{equation}

We next show that these $\Nphi$ states are mutually orthogonal and have equal norms. This allows us to construct them directly from the ED zero modes, without evaluating the integral in Eq.~\eqref{eq:v-definition} numerically.

\subsection{Construction of the coefficient states from ED}
\label{subsec:kernel-ED}

Consider the two electronic COM magnetic translations corresponding, in the rotated torus coordinates, to
\begin{equation}
\bm d_{\rm sh}=\frac{\bm C_2}{\Nphi},
\qquad
\bm d_{\rm cl}=-\frac{\bm C_1}{\Nphi}.
\end{equation}
We denote these translations by $\mathcal U_{\rm sh}$ and $\mathcal U_{\rm cl}$, respectively. Both preserve the uniform $V_1$ zero-mode space.

Applying these translations to the theta-function quasihole wavefunction in Eq.~\eqref{eq:qh-torus} gives their action on the coefficient states in Eq.~\eqref{eq:kernel-expand}. One translation shifts the scalar orbital index, while the other multiplies each orbital by an index-dependent phase:
\begin{equation}
\begin{split}
\mathcal U_{\rm cl}|v_j\rangle
&=
\beta_{\rm cl}
e^{\ii(2\pi j+\alpha)/\Nphi}|v_j\rangle,\\
\mathcal U_{\rm sh}|v_j\rangle
&=
\beta_{\rm sh}|v_{j+1}\rangle,
\qquad j=0,\ldots,\Nphi-2,
\end{split}
\label{eq:v-translations}
\end{equation}
with the final cyclic relation
\begin{equation}
\mathcal U_{\rm sh}|v_{\Nphi-1}\rangle
=
\beta_{\rm sh}e^{-\ii\beta}|v_0\rangle.
\label{eq:v-wrap}
\end{equation}
The phases $\beta_{\rm sh}$ and $\beta_{\rm cl}$ are fixed by the theta-function boundary conditions and the convention used for the electronic magnetic translations. They must be retained when changing the electron boundary twists.

The eigenvalues of $\mathcal U_{\rm cl}$ in Eq.~\eqref{eq:v-translations} are distinct. Since this operator is unitary, its eigenstates with different eigenvalues are orthogonal,
\begin{equation}
\langle v_i|v_j\rangle=0,\qquad i\ne j.
\end{equation}
Furthermore, $\mathcal U_{\rm sh}$ is unitary and maps each $|v_j\rangle$ to the next one. Therefore all their norms are equal:
\begin{equation}
\langle v_{j+1}|v_{j+1}\rangle
=
\langle v_j|v_j\rangle.
\end{equation}
Combining these results gives
\begin{equation}
\boxed{
\langle v_i|v_j\rangle=c\,\delta_{ij},
\qquad c>0.
}
\label{eq:isometry}
\end{equation}
We can consequently define the orthonormal many-electron states $|w_j\rangle=|v_j\rangle/\sqrt c$, giving
\begin{equation}
|\psi_\sigma^0(\eta)\rangle
=
\sqrt c\sum_{j=0}^{\Nphi-1}
\phi_j^{\alpha,\beta}(\eta)|w_j\rangle,
\qquad
\langle w_i|w_j\rangle=\delta_{ij}.
\label{eq:normalized-kernel}
\end{equation}

We now construct $|w_j\rangle$ from ED. Let $|e_\ell\rangle$, $\ell=0,\ldots,\Nphi-1$, be an orthonormal basis of the complete uniform $V_1$ zero-mode space, obtained by diagonalizing the parent Hamiltonian in its electronic COM sectors. Within this space the two magnetic translations are represented by the $\Nphi\times\Nphi$ matrices
\begin{equation}
[\widetilde{\mathcal U}_\mu]_{\ell\ell'}
=
\langle e_\ell|\mathcal U_\mu
|e_{\ell'}\rangle,
\qquad \mu={\rm sh},{\rm cl}.
\label{eq:projected-translations}
\end{equation}

These matrix elements can be calculated directly in the electronic Slater-determinant basis. The corresponding one-body magnetic translations permute LLL orbitals and multiply them by phases,
\begin{equation}
\mathcal U_\mu c_p^\dagger
\mathcal U_\mu^{-1}
=
e^{\ii\xi_{\mu,p}}
c_{\pi_\mu(p)}^\dagger.
\end{equation}
For a Slater determinant with occupied orbitals $I=(p_1,\ldots,p_{\Ne})$, we therefore have
\begin{equation}
\begin{split}
\mathcal U_\mu|I\rangle
={}&
\exp\!\left[
\ii\sum_{p\in I}\xi_{\mu,p}
\right]
\operatorname{sgn}(\pi_\mu|_I)\\
&\times|\pi_\mu(I)\rangle,
\end{split}
\label{eq:slater-translations}
\end{equation}
where the sign follows from reordering the transformed occupied orbitals into the fixed Slater-determinant ordering. The projected translations in Eq.~\eqref{eq:projected-translations} can therefore be constructed without forming dense matrices in the complete many-electron Hilbert space.

To obtain $|w_0\rangle$, we diagonalize $\widetilde{\mathcal U}_{\rm cl}$ and select the normalized eigenvector with the $j=0$ eigenvalue in Eq.~\eqref{eq:v-translations},
\begin{equation}
\mathcal U_{\rm cl}|w_0\rangle
=
\beta_{\rm cl}e^{\ii\alpha/\Nphi}|w_0\rangle.
\label{eq:clock-anchor}
\end{equation}
The remaining states are then obtained by repeatedly applying the shift translation,
\begin{equation}
\boxed{
|w_j\rangle
=
\beta_{\rm sh}^{-j}
\mathcal U_{\rm sh}^{\,j}|w_0\rangle,
\qquad j=1,\ldots,\Nphi-1.
}
\label{eq:shift-construction}
\end{equation}
The final application of $\mathcal U_{\rm sh}$ must reproduce the cyclic phase in Eq.~\eqref{eq:v-wrap}. This provides an independent check of the theta-function boundary conditions and fixes the correspondence between the ED zero modes and the scalar LLL orbitals, up to an overall phase.

Thus the states $|w_j\rangle$ are obtained entirely from the ED $V_1$ pseudopotential zero modes and their magnetic translations. No numerical evaluation of the many-electron wavefunction at different quasihole positions is required.

\subsection{Analytic matrix elements of the quasihole-coordinate phase}
\label{subsec:phase-matrices}

We next evaluate the operator $e^{\ii\bq\cdot\be}$ appearing in Eq.~\eqref{eq:anyon-link}. The quasihole-coordinate amplitude in Eq.~\eqref{eq:ed-coordinate} is obtained by taking the electronic bra of the localized quasihole state. Therefore, the relevant scalar basis consists of the complex-conjugate orbitals $[\phi_j^\lambda(\eta)]^*$, where $\lambda=(\alpha,\beta)$ denotes the covering-torus boundary phases.

At two neighboring momenta, the electron boundary twists, and hence $\lambda$, may be different. We write $\lambda'=(\alpha',\beta')$ for the boundary phases at the neighboring point. The required matrix elements are
\begin{equation}
\boxed{
M^{(-)}_{j'j}(\bq;\lambda',\lambda)
=
\int_{\mathcal C}d^2\eta\,
\phi_{j'}^{\lambda'}(\eta)
e^{\ii\bq\cdot\be}
[\phi_j^\lambda(\eta)]^*.
}
\label{eq:M-analytic-def}
\end{equation}
The superscript $(-)$ distinguishes the complex-conjugate scalar LLL basis from the holomorphic basis $\phi_j$.

It is useful to evaluate these matrix elements from the corresponding holomorphic LLL form factor,
\begin{equation}
M^{(+)}_{j'j}(\bq;\lambda',\lambda)
=
\int_{\mathcal C}d^2\eta\,
[\phi_{j'}^{\lambda'}(\eta)]^*
e^{\ii\bq\cdot\be}
\phi_j^\lambda(\eta).
\end{equation}
The two are related exactly by
\begin{equation}
\boxed{
M^{(-)}(\bq;\lambda',\lambda)
=
\left[
M^{(+)}(-\bq;\lambda',\lambda)
\right]^*.
}
\label{eq:M-conjugate}
\end{equation}

For the complex-conjugate basis, boundary matching requires
\begin{equation}
r=
\frac{-\bq\cdot\bm C_1-\alpha'+\alpha}{2\pi}
\in\mathbb Z,
\qquad
s=
\frac{-\bq\cdot\bm C_2-\beta'+\beta}{2\pi}
\in\mathbb Z.
\label{eq:M-selection}
\end{equation}
The integral along the first torus direction then requires $j'=j+r\pmod{\Nphi}$. Writing $t=(j+r-j')/\Nphi\in\mathbb Z$, the remaining Gaussian integral gives
\begin{equation}
\boxed{
\begin{aligned}
M^{(-)}_{j'j}(\bq;\lambda',\lambda)
={}&
\delta_{j',\,j+r\ ({\rm mod}\ \Nphi)}
\exp\!\left[
-\frac{\qell^2|\bq|^2}{4}
-\ii t\beta'\right.\\
&\left.
-\ii\qell^2
\left(q_y+\frac{V}{Y}q_x\right)
\left(p_j-\frac{q_x}{2}\right)
\right].
\end{aligned}
}
\label{eq:M-oblique}
\end{equation}
Here $p_j=(2\pi j+\alpha)/X$. The Gaussian depends on the Cartesian $|\bq|^2$. The term proportional to $V/Y$ comes from the tilted-torus phase in Eq.~\eqref{eq:oblique-orbitals}, while $e^{-\ii t\beta'}$ is the orbital-wrap phase of the conjugate coordinate basis.

Equation~\eqref{eq:M-oblique} provides the required matrix elements of $e^{\ii\bq\cdot\be}$ analytically. In particular, it retains the different boundary phases at the two endpoints and does not require numerical integration over the quasihole coordinate.

\subsection{Berry links in the uniform LLL}
\label{subsec:ED-link}

We first establish how the Berry links can be calculated from electronic states in the uniform LLL $V_1$ zero-mode space. The result holds for any state in this space and provides the starting point for the modulated ideal-band calculation below.

Recall that the orthonormal many-electron states $|w_j\rangle$ constructed in Sec.~\ref{subsec:kernel-ED} span the complete uniform one-quasihole zero-mode space. An arbitrary normalized state in this space can be expanded as
\begin{equation}
|\Phi_{\bk,a}\rangle
=
\sum_{j=0}^{\Nphi-1}
b_{j,a}(\bk)|w_j\rangle,
\qquad
b_{j,a}(\bk)
=
\langle w_j|\Phi_{\bk,a}\rangle.
\label{eq:electron-coefficients}
\end{equation}
The coefficients $b_{j,a}(\bk)$ can be obtained from the Slater-determinant coefficients of the electronic state using the previously constructed $|w_j\rangle$ basis. In the uniform LLL, the two folded quasihole states are degenerate, so the index $a$ simply labels a chosen basis within this degenerate space.

Using Eq.~\eqref{eq:normalized-kernel}, the quasihole-coordinate amplitude defined in Eq.~\eqref{eq:ed-coordinate} becomes
\begin{equation}
\begin{split}
\mathcal A_{a,\bk}(\eta)
&=
\langle\psi_\sigma^0(\eta)|
\Phi_{\bk,a}\rangle_e\\
&=
\sqrt c\sum_{j=0}^{\Nphi-1}
b_{j,a}(\bk)
[\phi_j^\lambda(\eta)]^*.
\end{split}
\label{eq:ED-coefficients}
\end{equation}
Since the electronic state is normalized and the $|w_j\rangle$ are orthonormal,
\begin{equation}
\sum_j|b_{j,a}(\bk)|^2=1.
\end{equation}
Together with the orthonormality of the scalar LLL orbitals $\phi_j^\lambda$, this gives
\begin{equation}
\int_{\mathcal C}d^2\eta\,
|\mathcal A_{a,\bk}(\eta)|^2=c.
\end{equation}
The normalized quasihole- coordinate wavefunction is therefore
\begin{equation}
\boxed{
\varphi_{a,\bk}(\eta)
=
\sum_{j=0}^{\Nphi-1}
b_{j,a}(\bk)
[\phi_j^\lambda(\eta)]^*.
}
\label{eq:coordinate-expansion}
\end{equation}

An important consequence is that this correspondence preserves the complete inner product. For any two electronic states in the uniform zero-mode space,
\begin{equation}
\begin{split}
&\int_{\mathcal C}d^2\eta\,
\varphi_{a,\bk}^*(\eta)
\varphi_{b,\bk}(\eta)\\
&\qquad =
\langle\Phi_{\bk,a}|
\Phi_{\bk,b}\rangle_e.
\end{split}
\end{equation}
Thus orthonormal electronic states correspond to orthonormal quasihole-coordinate wavefunctions. No additional orthonormalization is needed in the uniform LLL.

We can now evaluate the Berry links defined in Eq.~\eqref{eq:anyon-link}. Substituting Eq.~\eqref{eq:coordinate-expansion} and using the analytic matrix elements in Eq.~\eqref{eq:M-analytic-def}, we obtain
\begin{equation}
\boxed{
\mathcal L_{ab}(\bk',\bk)
=
\bm b_a(\bk')^\dagger
M^{(-)}(\bk'-\bk;\lambda',\lambda)
\bm b_b(\bk).
}
\label{eq:scalar-ED-link}
\end{equation}
Here $\bm b_a(\bk)$ denotes the column vector with components $b_{j,a}(\bk)$. The matrix $M^{(-)}$ accounts for the insertion $e^{\ii(\bk'-\bk)\cdot\be}$ and the different covering-torus boundary phases at the two momenta. Equation~\eqref{eq:scalar-ED-link} therefore gives the quasihole-coordinate Berry links directly from the electronic coefficients, without numerical integration over $\eta$.

For a folded quasihole LLL, the overlap in Eq.~\eqref{eq:scalar-ED-link} is a $2\times2$ matrix. Since the two states are degenerate in the uniform LLL problem, their individual Chern numbers are not determined by the Hamiltonian. We instead calculate the Chern number of the combined two-dimensional space using the non-Abelian Berry links.

The uniform result establishes the exact correspondence between electronic zero modes and quasihole-coordinate wavefunctions. In the modulated ideal band, the same construction applies, but the electronic states and their coordinate-space inner product are modified. We now derive the corresponding Berry links for the screened-Coulomb quasihole bands.

\subsection{Berry links in the modulated ideal Chern band}

We now apply the construction to the modulated ideal Chern bands considered in the main text. As established in Sec.~\ref{sec:folding}, multiplication of the many-electron wavefunction by $\prod_i h(\bm r_i)$ is represented in the orthonormal electronic occupation bases by the invertible map $R_h$. The localized quasihole state becomes
\begin{equation}
|\psi_\sigma^h(\eta)\rangle
=
g(\eta)R_h|\psi_\sigma^0(\eta)\rangle.
\end{equation}
Using Eq.~\eqref{eq:normalized-kernel}, we obtain
\begin{equation}
|\psi_\sigma^h(\eta)\rangle
=
\sqrt c\,g(\eta)
\sum_j\phi_j^\lambda(\eta)
R_h|w_j\rangle.
\label{eq:modulated-kernel}
\end{equation}

Unlike in the uniform LLL, the two quasihole bands can now acquire different dispersions. The electronic states whose topology we calculate are the eigenstates $|\Phi_{\bk,\pm}\rangle$ defined in Eq.~\eqref{eq:anyon-eigenstates}, obtained by diagonalizing the screened Coulomb interaction in the two-dimensional modulated $V_1$ zero-mode space. Thus the following Berry links are constructed from the same eigenstates whose energies are plotted in the main text.

The quasihole-coordinate amplitude of such an eigenstate is
\begin{equation}
\begin{split}
\mathcal A^h_{a,\bk}(\eta)
&=
\langle\psi_\sigma^h(\eta)|
\Phi_{\bk,a}\rangle_e\\
&=
\sqrt c\,g^*(\eta)
\sum_j a^h_{j,a}(\bk)
[\phi_j^\lambda(\eta)]^*,
\end{split}
\label{eq:modulated-amplitude}
\end{equation}
where
\begin{equation}
\boxed{
a^h_{j,a}(\bk)
=
\langle R_hw_j|\Phi_{\bk,a}\rangle_e
=
\langle w_j|R_h^\dagger
|\Phi_{\bk,a}\rangle_e.
}
\label{eq:modulated-coefficients}
\end{equation}
These coefficients are obtained from the screened-Coulomb eigenstates and the known many-body map $R_h$. Because $R_h$ is not unitary, their ordinary Euclidean norm need not be one.

The second difference from the uniform LLL is the factor $g^*(\eta)$ in Eq.~\eqref{eq:modulated-amplitude}. Its contribution to the coordinate-space overlaps is determined by $|g(\eta)|^2$. Expanding this function in reciprocal harmonics of the physical moir\'e lattice,
\begin{equation}
|g(\eta)|^2
=
\sum_{\bG}v_{\bG}
e^{\ii\bG\cdot\be},
\end{equation}
we obtain the matrix elements
\begin{equation}
\boxed{
M_g^{(-)}(\bq;\lambda',\lambda)
=
\sum_{\bG}v_{\bG}
M^{(-)}(\bq+\bG;\lambda',\lambda).
}
\label{eq:g-convolution}
\end{equation}
Thus the effect of the modulation is included through a sum of the analytic uniform-LLL matrix elements derived in Sec.~\ref{subsec:phase-matrices}.

For an isolated quasihole band, we first normalize its coordinate wavefunction. Defining
\begin{equation}
\mathcal N_a(\bk)
=
[\bm a_a^h(\bk)]^\dagger
M_g^{(-)}(0;\lambda,\lambda)
\bm a_a^h(\bk),
\end{equation}
where $\bm a_a^h(\bk)$ is the column vector with components $a^h_{j,a}(\bk)$, we have
\begin{equation}
\varphi_{a,\bk}(\eta)
=
\frac{\mathcal A^h_{a,\bk}(\eta)}
{\sqrt{c\,\mathcal N_a(\bk)}}.
\end{equation}
Substituting this expression into Eq.~\eqref{eq:anyon-link} gives
\begin{equation}
\boxed{
\mathcal L_g(\bk',\bk)
=
\frac{
[\bm a_a^h(\bk')]^\dagger
M_g^{(-)}(\bk'-\bk;\lambda',\lambda)
\bm a_a^h(\bk)
}{
\sqrt{\mathcal N_a(\bk')}
\sqrt{\mathcal N_a(\bk)}
}.
}
\label{eq:g-link-scalar}
\end{equation}
This is the Berry-link formula used for the isolated quasihole bands in the main figures. The electronic eigenstates enter through $\bm a_a^h(\bk)$, while $M_g^{(-)}$ incorporates the quasihole coordinate and the ideal-band modulation.

When the two quasihole bands touch, their individual Berry curvatures are not globally defined. We must then retain the complete two-dimensional subspace. More generally, for $r$ bands, we collect their coefficient vectors into the $\Nphi\times r$ matrix
\begin{equation}
A(\bk)
=
\big(
\bm a_1^h(\bk),\ldots,
\bm a_r^h(\bk)
\big).
\end{equation}
The coordinate-space overlap matrix at the same momentum is
\begin{equation}
\mathcal G(\bk)
=
A(\bk)^\dagger
M_g^{(-)}(0;\lambda,\lambda)
A(\bk).
\end{equation}
Using its positive inverse square root, the normalized matrix of Berry links is
\begin{equation}
\boxed{
\begin{aligned}
\mathcal L_g(\bk',\bk)
={}&
\mathcal G(\bk')^{-1/2}
A(\bk')^\dagger\\
&\times
M_g^{(-)}(\bk'-\bk;\lambda',\lambda)
A(\bk)
\mathcal G(\bk)^{-1/2}.
\end{aligned}
}
\label{eq:g-link}
\end{equation}
For $r=1$, this reduces to Eq.~\eqref{eq:g-link-scalar}. For the two connected bands, $r=2$, it gives the non-Abelian Berry links used to calculate the Chern number of the complete subspace.

We finally clarify why the normalization requires additional care in the modulated system. The electronic eigenstates are orthonormal,
\begin{equation}
\langle\Phi_{\bk,a}|
\Phi_{\bk,b}\rangle_e
=
\delta_{ab}.
\end{equation}
In the uniform LLL, their coordinate wavefunctions $\varphi_{a,\bk}(\eta)$ inherit this orthogonality exactly. In the modulated ideal band, however, $R_h$ is nonunitary and the coordinate wavefunctions contain the additional factor $g^*(\eta)$. Consequently, their coordinate-space inner products need not equal the electronic ones:
\begin{equation}
\int_{\mathcal C}d^2\eta\,
\varphi_{a,\bk}^*(\eta)
\varphi_{b,\bk}(\eta)
\neq
\delta_{ab}
\end{equation}
in general. Normalizing each wavefunction individually fixes its norm, but does not necessarily make different wavefunctions orthogonal.

The complete overlap matrix $\mathcal G(\bk)$ accounts for this difference. In particular, the coefficient matrix $A(\bk)\mathcal G(\bk)^{-1/2}$ defines mutually orthonormal quasihole-coordinate wavefunctions. For the complete two-band space, this is the inner-product-preserving representation of the electronic states. The orthonormalization is a change of representation; it does not alter the screened-Coulomb eigenvalues.

We have independently checked the effect of this orthonormalization on the Berry curvature. For the isolated bands, the local curvature obtained from the individually normalized wavefunctions differs slightly from that obtained using the inner-product-preserving representation, but the integrated Chern numbers are identical. For the connected two-band space, the full-Gram normalization in Eq.~\eqref{eq:g-link} already implements this orthonormalization, and the trace Berry curvature is unchanged.

Therefore, Eqs.~\eqref{eq:g-link-scalar} and \eqref{eq:g-link} provide the Berry links used to characterize the screened-Coulomb quasihole bands. The former applies to an isolated band; the latter applies to the complete subspace when the bands are connected. The resulting Chern numbers are unchanged by the normalization check, although the local Berry curvature depends slightly on the chosen coordinate-space representation.

\subsection{Lattice Berry flux and the uniform LLL Chern number}
\label{subsec:uniform-curvature}

We calculate the Chern number from the quasihole-coordinate links constructed above. For an isolated band, define the unit-modulus link
\begin{equation}
\mathcal U_\mu(\bk)
=
\frac{
\mathcal L_g(\bk+\bq_\mu,\bk)
}{
|\mathcal L_g(\bk+\bq_\mu,\bk)|
}.
\end{equation}
For a rank-$r$ subspace, we retain the full $r\times r$ overlap matrix and take the unitary part of its polar decomposition,
\begin{equation}
\mathcal L_g=V\Sigma W^\dagger,
\qquad
\mathcal U=VW^\dagger.
\end{equation}
With links directed from $\bk$ to $\bk+\bq_\mu$, the Wilson plaquette is
\begin{equation}
\begin{split}
\mathcal W(\bk)
={}&
\mathcal U_2(\bk)^\dagger
\mathcal U_1(\bk+\bq_2)^\dagger\\
&\times
\mathcal U_2(\bk+\bq_1)
\mathcal U_1(\bk).
\end{split}
\end{equation}
The Berry flux through the plaquette and the Chern number are
\begin{equation}
\boxed{
\gamma(\bk)
=
\Arg\det\mathcal W(\bk),
\qquad
C=
\frac1{2\pi}
\sum_{\bk\in{\rm aBZ}}\gamma(\bk).
}
\label{eq:Wilson-plaquette}
\end{equation}
The determinant phase is invariant under changes of basis within the rank-$r$ subspace. At a magnetic-BZ boundary, the momentum transfer must be the short unwrapped displacement, with the endpoint identified through the appropriate magnetic-translation boundary condition. In the present coordinate representation, these boundary phases are the complex conjugates of those of the holomorphic scalar LLL basis.

We now calculate the quasihole Berry curvature analytically in the uniform LLL. This also explains why a scalar quasihole-coordinate wavefunction can be used to characterize the topology of an underlying correlated many-electron state.

Recall that the localized uniform Laughlin quasihole wavefunction admits the exact expansion
\begin{equation}
|\psi_\sigma^0(\eta)\rangle
=
\sqrt{c}\sum_j
\phi_j(\eta)|w_j\rangle,
\end{equation}
where the $|w_j\rangle$ form an orthonormal basis of the many-electron one-quasihole zero-mode space, and $\phi_j(\eta)$ are orthonormal orbitals of a scalar LLL with effective magnetic length $\qell^2=m\ell_B^2$. The constant $c$ is independent of $j$, as established from magnetic translation symmetry.

An arbitrary normalized electronic state in this zero-mode subspace can be expanded as
\begin{equation}
|\Phi_{\bk}\rangle
=
\sum_j b_j(\bk)|w_j\rangle,
\qquad
\sum_j|b_j(\bk)|^2=1.
\end{equation}
Its normalized quasihole-coordinate wavefunction is therefore
\begin{equation}
\begin{split}
\varphi_{\bk}(\eta)
&\equiv
\frac1{\sqrt c}
\langle\psi_\sigma^0(\eta)|
\Phi_{\bk}\rangle_e\\
&=
\sum_j b_j(\bk)[\phi_j(\eta)]^*.
\end{split}
\end{equation}
Thus integrating out the electronic coordinates maps the many-electron zero-mode space exactly onto the complex-conjugate scalar LLL. Since both sets of basis states are orthonormal, this correspondence preserves inner products. In particular, $\varphi_{\bk}(\eta)$ contains the full information about $|\Phi_{\bk}\rangle$ within the zero-mode subspace.

The magnetic translation eigenvalues are also preserved under this correspondence, with the momentum convention established above. Consequently, an electronic zero-mode state at momentum $\bk$ maps to a scalar LLL magnetic Bloch state at the same momentum. In the unfolded uniform problem, there is one such state at each momentum when the magnetic unit cell contains one effective flux quantum. The coordinate wavefunction $\varphi_{\bk}(\eta)$ must therefore coincide, up to a momentum-dependent phase, with the corresponding scalar LLL Bloch orbital.

To evaluate its Berry curvature explicitly, consider a rectangular quasihole magnetic unit cell of lengths $a$ and $b$, satisfying
\begin{equation}
ab=2\pi\qell^2,
\qquad
\qell^2=m\ell_B^2.
\end{equation}
We choose the following normalized Bloch wavefunction in the conventional quasihole-coordinate representation:
\begin{equation}
\begin{split}
\varphi_{\bk}(x,y)
={}&
\frac{e^{-\ii xy/(2\qell^2)}}
{\sqrt{a\sqrt{\pi\qell^2}}}
\sum_{n\in\mathbb Z}
e^{\ii n b k_y}
e^{\ii(k_x+2\pi n/a)x}\\
&\times
\exp\!\left[
-\frac{
[y-\qell^2(k_x+2\pi n/a)]^2
}{2\qell^2}
\right],
\qquad
u_{\bk}(\eta)
=
e^{-\ii\bk\cdot\be}\varphi_{\bk}(\eta).
\end{split}
\label{eq:uniform-wavefunction}
\end{equation}
Here $\eta=x+\ii y$, and the Gaussian centers are located at $y_n=\qell^2 k_x+nb$, where we have used $ab=2\pi\qell^2$. This Bloch wavefunction belongs to the complex-conjugate scalar LLL obtained from the many-electron quasihole state by the projection above.

The Berry connection of the cell-periodic coordinate wavefunction is
\begin{equation}
\mathscr A_\mu(\bk)
=
\ii\int_{\mathcal C}d^2\eta\,
u_{\bk}^*(\eta)
\partial_{k_\mu}u_{\bk}(\eta),
\label{eq:coordinate-Berry}
\end{equation}
where the integral in this analytic Bloch calculation is over one magnetic unit cell.

After removing the Bloch phase, the $k_x$ dependence of $u_{\bk}$ enters only through the real Gaussian factors. Integration over one period in $x$ eliminates cross terms between different $n$, so the remaining contribution to $\langle u_{\bk}|\partial_{k_x}u_{\bk}\rangle$ is one-half the derivative of the normalization. Since the wavefunction is normalized for every $\bk$, this gives
\begin{equation}
\mathscr A_x(\bk)=0.
\end{equation}

For the $k_y$ component, the derivative acts on $e^{\ii n b k_y}e^{-\ii k_y y}$, giving $\ii(nb-y)$. Consequently,
\begin{equation}
\mathscr A_y(\bk)
=
\langle y-nb\rangle,
\end{equation}
where the expectation value includes the sum over Gaussian images. Because the Gaussian in the $n$th term is centered at $y_n=\qell^2 k_x+nb$, and the intervals obtained by shifting $y$ by $nb$ cover the real line, we have
\begin{equation}
\langle y-nb\rangle
=
\qell^2 k_x.
\end{equation}
We therefore obtain
\begin{equation}
\boxed{
\mathscr A_x=0,\qquad
\mathscr A_y=\qell^2 k_x,\qquad
F_{xy}^{\rm LLL}
=
\partial_{k_x}\mathscr A_y
-
\partial_{k_y}\mathscr A_x
=
\qell^2=m\ell_B^2.
}
\label{eq:uniform-F}
\end{equation}

The unfolded magnetic anyon BZ has area $(2\pi/a)(2\pi/b)=2\pi/\qell^2$. The corresponding Chern number is therefore
\begin{equation}
\boxed{
C_{\rm LLL}
=
\frac1{2\pi}
\int_{\rm aBZ}d^2k\,
F_{xy}^{\rm LLL}
=
+1.
}
\label{eq:uniform-Chern}
\end{equation}
A momentum-dependent phase relating the ED-derived coordinate wavefunction to the Bloch orbital chosen above changes the Berry connection by a gradient, but leaves the curvature and Chern number unchanged. Thus $C_{\rm LLL}=+1$ applies directly to the coordinate wavefunction obtained from $\langle\psi_\sigma^0(\eta)|
\Phi_{\bk}\rangle_e$.

For a two-flux moir\'e cell, the uniform quasihole band folds into two degenerate states at each momentum of a magnetic BZ with half the original area. The trace curvature of this rank-two space is
\begin{equation}
\boxed{
\operatorname{tr}F_{xy}^{\rm folded,LLL}
=
2m\ell_B^2,
\qquad
\frac1{2\pi}
\int_{{\rm aBZ}_{\rm fold}}d^2k\,
\operatorname{tr}F_{xy}^{\rm folded,LLL}
=
+1.
}
\label{eq:folded-uniform-F}
\end{equation}
For $m=3$, the uniform trace curvature is therefore $6\ell_B^2$. Since the two folded states are degenerate in the uniform LLL, their individual Chern numbers are not determined by the energy spectrum. Only the Chern number of their combined space is unambiguously defined.

When the screened interaction splits the two quasihole bands in the modulated system, the corresponding electronic eigenstates $|\Phi_{\bk,\pm}\rangle$ define separate bands, each of which can carry its own Chern number. These integers need not equal the Chern number of the parent quasihole LLL, although their sum remains $+1$ as long as the complete two-band subspace remains smoothly connected to the folded parent space. When crystalline symmetry enforces band crossings, the individual Chern numbers are not defined, but the combined rank-two Chern number remains well defined.

\subsection{Physical meaning of the quasihole-band Chern number}
\label{subsec:qh-chern-physical}

We now explain why the Berry curvature calculated from $\varphi_{a,\bk}(\eta)$ characterizes the effective motion of a Laughlin quasihole. The essential point is that $\eta$ represents the position of the fractional excitation, and that a slowly varying electric potential acts on this coordinate through the fractional quasihole charge.

The starting point is the exact correspondence established above for the uniform LLL. Equation~\eqref{eq:normalized-kernel} gives
\begin{equation}
\begin{split}
&\langle\psi_\sigma^0(\eta')|
\psi_\sigma^0(\eta)\rangle_e\\
&\qquad =
c\sum_j
[\phi_j^\lambda(\eta')]^*
\phi_j^\lambda(\eta).
\end{split}
\label{eq:qh-coherent-overlap}
\end{equation}
Thus the overlap between two localized many-electron quasihole states is determined exactly by that of scalar LLL states with magnetic length $\qell=\sqrt m\ell_B$. The physical significance of this correspondence follows from the Berry phase acquired by adiabatically transporting a localized Laughlin quasihole around a closed path. This phase is that of a particle carrying fractional charge $e/m$ in the uniform magnetic field ~\cite{arovas1984fractional}. The corresponding effective magnetic length is precisely $\qell^2=m\ell_B^2$, in agreement with the scalar LLL representation of $\varphi_{a,\bk}(\eta)$. On the torus, the resulting quantum motion is described by the quasihole magnetic translation algebra derived earlier.

The correspondence between the electronic zero modes and the quasihole-coordinate wavefunctions $\varphi_{a,\bk}(\eta)$ is exact at fixed electronic boundary twists. In particular, Eq.~\eqref{eq:normalized-kernel} gives the resolution of the projector onto the uniform one-quasihole zero-mode space,
\begin{equation}
\boxed{
P_0
=
\frac1c
\int_{\mathcal C}d^2\eta\,
|\psi_\sigma^0(\eta)\rangle
\langle\psi_\sigma^0(\eta)|.
}
\label{eq:qh-resolution}
\end{equation}
An arbitrary state $|\Phi\rangle$ in this space can therefore be reconstructed from its quasihole wavefunction as
\begin{equation}
|\Phi\rangle_e
=
\frac1{\sqrt c}
\int_{\mathcal C}d^2\eta\,
|\psi_\sigma^0(\eta)\rangle
\varphi(\eta).
\label{eq:qh-reconstruction}
\end{equation}
Consequently, any Hamiltonian projected into this electronic zero-mode space has an exact representation acting on $\varphi(\eta)$. For the screened interaction $H_{\rm eff}=P_0H_{\rm sc}P_0$, the Schr\"odinger equation becomes
\begin{equation}
\begin{split}
\ii\hbar\partial_t\varphi(\eta,t)
={}&
\frac1c
\int_{\mathcal C}d^2\eta'\,
\langle\psi_\sigma^0(\eta)|
H_{\rm eff}
|\psi_\sigma^0(\eta')\rangle_e\\
&\times\varphi(\eta',t).
\end{split}
\label{eq:qh-effective-schrodinger}
\end{equation}
Thus the projected electronic Hamiltonian and its quasihole representation have identical spectra and dynamics within the zero-mode space. The corresponding construction in the modulated ideal band therefore describes the same screened-Coulomb eigenstates $|\Phi_{\bk,\pm}\rangle$ whose dispersions are calculated in the main text.

To connect this quasihole description to a physical electric perturbation, we first define the quasihole number-density operator. In the uniform one-quasihole zero-mode space, let
\begin{equation}
\widehat\rho_{\rm qh}(\eta)
=
\frac1c
|\psi_\sigma^0(\eta)\rangle
\langle\psi_\sigma^0(\eta)|.
\end{equation}
For a normalized electronic state $|\Phi\rangle$,
\begin{equation}
\langle\Phi|
\widehat\rho_{\rm qh}(\eta)
|\Phi\rangle_e
=
|\varphi(\eta)|^2,
\end{equation}
which gives the quasihole probability density in this representation. Its Fourier component is
\begin{equation}
\boxed{
\widehat\rho_{\rm qh}(\bq)
=
\frac1c
\int_{\mathcal C}d^2\eta\,
|\psi_\sigma^0(\eta)\rangle
e^{-\ii\bq\cdot\be}
\langle\psi_\sigma^0(\eta)|.
}
\label{eq:qh-position-operator}
\end{equation}
It satisfies $\widehat\rho_{\rm qh}(0)=P_0$, corresponding to one quasihole, and its matrix elements in the orthonormal $|w_j\rangle$ basis are
\begin{equation}
\langle w_{j'}|
\widehat\rho_{\rm qh}(\bq)
|w_j\rangle_e
=
M^{(-)}_{j'j}(-\bq;\lambda,\lambda).
\end{equation}
Thus the matrices $M^{(-)}$ used in our Berry-link calculation are also the matrix elements of the quasihole density operator in its effective LLL.

Under the exact correspondence with the scalar quasihole LLL, the projected density insertion is
\begin{equation}
P_{\rm LLL}^{\eta}
e^{-\ii\bq\cdot\be}
P_{\rm LLL}^{\eta}
=
e^{-\qell^2|\bq|^2/4}
e^{-\ii\bq\cdot\widehat{\mathbf R}_{\rm qh}},
\label{eq:qh-guiding-insertion}
\end{equation}
where $\widehat{\mathbf R}_{\rm qh}$ denotes the quasihole guiding-center position operator. The Gaussian is the LLL form factor, while the remaining factor is the guiding-center density operator. This identifies the quasihole density within the effective LLL, but it is distinct from the microscopic electronic density.

We now relate these two densities. The physical electronic number-density operator is
\begin{equation}
\rho_e(\bq)
=
\sum_{i=1}^{\Ne}
e^{-\ii\bq\cdot\bm r_i}.
\end{equation}
A localized Laughlin quasihole carries an electronic number deficit $-1/m$ relative to the incompressible background. For a quasihole centered at $\eta$, the Fourier transform of this density deficit therefore has the long-wavelength form
\begin{equation}
\delta n_e(\bq;\eta)
=
-\frac1m e^{-\ii\bq\cdot\be}
+O(|\bq|\xi_{\rm qh}),
\end{equation}
where $\xi_{\rm qh}$ characterizes the spatial extent of the quasihole charge deficit. The leading term follows from its total fractional charge, while the corrections describe the spatial distribution of the missing electronic charge.

This motivates the corresponding relation between the operators projected into the low-energy one-quasihole space,
\begin{equation}
\boxed{
P_0\delta\rho_e(\bq)P_0
=
-\frac1m\widehat\rho_{\rm qh}(\bq)
+
O(|\bq|\xi_{\rm qh}).
}
\label{eq:qh-density-longwave}
\end{equation}
Here $\delta\rho_e(\bq)$ is the electronic density relative to the appropriate incompressible background. The relation is understood at the level of low-energy matrix elements in the long-wavelength limit, assuming a localized quasihole in a gapped, screened Laughlin phase. For the modulated system, $P_0$ and $\widehat\rho_{\rm qh}$ denote the corresponding projected operators in its quasihole representation.

In the uniform LLL, continuous magnetic translation symmetry imposes the stronger relation
\begin{equation}
P_0\delta\rho_e(\bq)P_0
=
d(\bq)\widehat\rho_{\rm qh}(\bq)
\end{equation}
at every compatible momentum, where the finite-momentum form factor $d(\bq)$ approaches $-1/m$ at long wavelengths. In the modulated ideal band, continuous magnetic translation symmetry is absent. The projected electronic density can therefore contain additional matrix elements that are not proportional to the quasihole density, including contributions from the intracell charge distribution and mixing between quasihole bands. Our microscopic calculations confirm such corrections at finite momenta that decreases with at the increasing wavelengths. Equation~\eqref{eq:qh-density-longwave} is thus a long-wavelength relation, rather than an exact identity at arbitrary momentum.

The physical consequence is that a smooth external electric potential acts, to leading order, as a potential on the fractionally charged quasihole. Indeed, taking the electron charge to be $-e$ and writing $V_{\rm ext}(\bm r)$ for the electric scalar potential, the perturbation relative to the incompressible background is
\begin{equation}
\begin{split}
H_{\rm ext}
&=
-e\int d^2r\,
V_{\rm ext}(\bm r)\,
\delta\rho_e(\bm r)\\
&\simeq
\frac{e}{m}
\int d^2\eta\,
V_{\rm ext}(\be)\,
\widehat\rho_{\rm qh}(\eta).
\end{split}
\label{eq:qh-electric-coupling}
\end{equation}
The leading coupling is therefore that of a quasihole carrying charge $+e/m$. The finite-momentum corrections to Eq.~\eqref{eq:qh-density-longwave} give additional, nonuniversal contributions to the microscopic response.

Having established how an external potential acts on the quasihole, we can relate its motion to the quasihole-band Berry curvature. For an isolated band, the Berry connection of the cell-periodic quasihole Bloch wavefunction is
\begin{equation}
\mathscr A^\varphi_{a,\mu}(\bk)
=
\ii\langle u_{a,\bk}|
\partial_{k_\mu}u_{a,\bk}\rangle_\eta.
\end{equation}
The quasihole position projected into that band acts locally in momentum space as
\begin{equation}
\widehat R_{{\rm qh},\mu}^{(a)}
=
\ii\partial_{k_\mu}
+
\mathscr A^\varphi_{a,\mu}(\bk),
\label{eq:qh-band-position}
\end{equation}
with
\begin{equation}
\left[
\widehat R_{{\rm qh},x}^{(a)},
\widehat R_{{\rm qh},y}^{(a)}
\right]
=
\ii F^\varphi_{xy,a}(\bk).
\label{eq:qh-band-position-commutator}
\end{equation}
Thus, under a weak, slowly varying electric potential, the effective quasihole dynamics is governed by its band dispersion and Berry curvature. The latter contributes to the anomalous motion of a quasihole wavepacket, while its integral over the magnetic anyon BZ gives the Chern number of the isolated band. For symmetry-connected bands, the corresponding description involves the non-Abelian Berry connection of the complete subspace.

In the folded ideal Chern-band systems considered here, moir\'e periodicity produces two quasihole states at each momentum in the magnetic anyon BZ. The screened Coulomb interaction gives these states finite dispersion and can split them into isolated bands with distinct Chern numbers, while crystalline symmetries can protect crossings between them. Consequently, a physical electric potential acts on fractionally charged quasiholes whose effective motion is governed by band dispersions and topology that are absent in the uniform LLL. This establishes the physical significance of the quasihole-band Chern numbers calculated in this work.

\subsection{Distinction from the electronic boundary-twist Chern number}
\label{subsec:qh-versus-electron-chern}

We finally clarify why the Berry curvature of $\varphi_{a,\bk}(\eta)$ is used to characterize the quasihole bands, rather than the Berry curvature obtained by directly differentiating the electronic eigenstates $|\Phi_{\bk,a}\rangle$ with respect to boundary twists.

The electronic boundary-twist Chern number is obtained by following the many-electron states as magnetic flux is inserted through the cycles of the torus. After identifying the electronic Hilbert spaces at different twists in a consistent gauge, the corresponding Berry connection can be written schematically as
\begin{equation}
\mathscr A^e_{a,\mu}
=
\ii\langle\Phi_{\bk,a}|
\partial_{k_\mu}\Phi_{\bk,a}\rangle_e,
\end{equation}
where the momentum variation is understood through its relation to the electronic boundary twists. The common electronic gauge and boundary sewing are essential for this expression. This construction probes the many-electron response to flux insertion.

The quasihole-band Chern number, by contrast, is defined through $\varphi_{a,\bk}(\eta)$ and its magnetic Bloch boundary conditions. The coordinate $\eta$ represents the position of the fractionally charged quasihole, whose effective magnetic translation algebra follows from the microscopic Berry phase of moving the localized Laughlin quasihole. The resulting Berry curvature characterizes the geometry of the quasihole energy band, including its effective position algebra and the associated wavepacket dynamics.

The two Berry connections are not generally equal. As shown in Eq.~\eqref{eq:coordinate-derivative}, the momentum derivative of the quasihole wavefunction contains contributions from both the electronic eigenstate $|\Phi_{\bk,a}\rangle$ and the localized quasihole state $|\psi_\sigma(\eta)\rangle$. Furthermore, the electronic flux continuation and the quasihole magnetic BZ involve different global identifications. The exact correspondence between electronic zero modes and scalar quasihole wavefunctions at fixed twists therefore does not imply equality of their Berry curvatures when the twists are varied.

This distinction already appears in the uniform LLL. The quasihole Bloch-wavefunction construction gives $C_{\rm LLL}=+1$, as derived in Eq.~\eqref{eq:uniform-Chern}. To calculate the electronic boundary-twist Chern number instead, we consider the complete one-quasihole zero-mode space, which has dimension $\Nphi=m\Ne+1$. Let $|e_\ell(\bm\theta)\rangle$, $\ell=0,\ldots,\Nphi-1$, be an orthonormal basis of this space at electronic boundary twists $\bm\theta=(\theta_1,\theta_2)$. The Berry links are constructed from the full electronic overlap matrices
\begin{equation}
\mathcal L^e_{\ell\ell'}(\bm\theta',\bm\theta)
=
\langle e_\ell(\bm\theta')|
e_{\ell'}(\bm\theta)\rangle_e,
\label{eq:electronic-twist-link}
\end{equation}
after expressing the states at different twists in a common electronic gauge. In particular, the variation of the single-particle orbitals with the boundary twists and the electronic boundary sewing must be retained.

Calculating the non-Abelian Chern number of this complete zero-mode space over one electronic boundary-twist torus, $\theta_1,\theta_2\in[0,2\pi)$, gives
\begin{equation}
\boxed{
C_e
=
\frac{1}{2\pi}
\int_0^{2\pi}d\theta_1
\int_0^{2\pi}d\theta_2\,
\operatorname{tr}F^e_{\theta_1\theta_2}
=
-\Ne.
}
\label{eq:electronic-zero-mode-chern}
\end{equation}
This agrees, up to the opposite sign convention for the underlying LLL, with the result of Zeybek and Umucalilar~\cite{zeybek2022structure}, who calculated the many-body Chern number of quasihole manifolds under electronic flux insertion. For a single quasihole, their counting gives a Chern number of magnitude $\Ne$ for the complete $\Nphi$-dimensional manifold. Thus our direct electronic calculation reproduces the previously established result, while the quasihole Bloch wavefunction gives $C_{\rm LLL}=+1$.

These different integers do not indicate that the quasihole representation omits electronic states. Rather, the map between the electronic and quasihole representations itself depends on the boundary twists, and their global boundary identifications differ. In particular, $C_e=-\Ne$ is defined over one physical electronic boundary-twist torus, whereas $C_{\rm LLL}=+1$ is defined over the magnetic anyon BZ.

Both constructions have physical meaning, but they answer different questions. The electronic boundary-twist Chern number characterizes the many-body response to flux insertion. The quasihole-band Chern number characterizes the effective band structure in which the fractionally charged quasihole moves. The latter is the quantity relevant to the moir\'e-induced quasihole-band splitting and connectivity studied here.

We therefore use Eqs.~\eqref{eq:anyon-link} and \eqref{eq:Wilson-plaquette} to calculate the Chern numbers of the screened-Coulomb quasihole bands $|\Phi_{\bk,\pm}\rangle$. In the two-flux moir\'e problem, the combined quasihole Chern number is $+1$. When the bands are separated, their individual Chern numbers can differ from that of the parent quasihole LLL, while their sum remains $+1$. When crystalline symmetry protects band crossings, only the Chern number of the connected two-band subspace is defined.

\end{document}